\documentclass[aps,prl,reprint,superscriptaddress,letterpaper,twocolumn,amsmath,amssymb,nolongbibliography,10pt]{revtex4-2}

\usepackage{graphicx}% Include figure files
\usepackage{dcolumn}% Align table columns on decimal point
\usepackage{bm}% bold math
\usepackage{lipsum}
\usepackage{array,soul}
\usepackage{siunitx}
\usepackage{color}
\usepackage{lastpage,fancyhdr,graphicx}
\usepackage{mathrsfs}
\usepackage[hidelinks]{hyperref}
\newcommand{\av}[1]{\langle #1 \rangle}
\DeclareMathOperator*{\argmax}{arg\,max}

\usepackage[table,xcdraw]{xcolor}
\definecolor{mDarkRed}{HTML}{CF0A0A}

\newcommand{\ce}{\emph{C. elegans}}

\begin{document}

\title{The exploration-exploitation paradigm for networked biological systems}% Force line breaks with \\

\author{Vito Dichio}
\author{Fabrizio De Vico Fallani}%
\affiliation{Sorbonne Universite, Paris Brain Institute - ICM, CNRS, Inria, Inserm, AP-HP, Hopital de la Pitie Salpêtriere, F-75013, Paris, France}

\date{\today}

\begin{abstract}
The stochastic exploration of the configuration space and the exploitation of functional states underlie many biological processes. The evolutionary dynamics stands out as a remarkable example. Here, we introduce a novel formalism that mimics evolution and encodes a general exploration-exploitation dynamics for biological networks. We apply it to the brain wiring problem, focusing on the maturation of that of the nematode \ce. We demonstrate that a parsimonious maxent description of the adult brain combined with our framework is able to track down the entire developmental trajectory. 
\end{abstract}

\maketitle

\paragraph{Introduction.}

% General context
Modelling and analyzing the dynamics of biological systems is notoriously challenging. Critically, they are often stochastic in nature as they involve and possibly exploit some degree of randomness. At the same time, biological dynamics are also shaped by functional constraints that determine which outcomes are viable. The constraints emerge from the need for biological systems to perform specific tasks, and act on the system as a whole, not on specific components. Stochastic events that violate these constraints are unlikely to persist, while those that align with them are more likely to become integrated. If the details of such exploration-exploitation (EE) dynamics are context-dependent, general principles can still be formulated \cite{bialek2012}. This entails addressing several challenging questions. For instance, how do biological systems explore the space of possible configurations? How do they identify the optimal states that satisfy specific functional demands? 

% The sub-case of evolution
A possible solution is offered by Nature itself. In evolutionary dynamics, a population primarily evolves under the combined action of mutations and recombinations (exploration) and natural selection (exploitation). The latter is based on the notion of fitness: those individuals that are more apt to the environment will have a higher reproductive success (high fitness) and survive to the next generations, while the others will go extinct, Supplemental material Section I (SM-I) \cite{crow2017}. We argue that evolutionary dynamics is a particular instance of the aforementioned EE dynamics and build upon it to construct a general EE formalism for networked biological systems. 

We use it to tackle the brain wiring problem and model the developmental dynamics of the \ce\ connectome, recently obtained by serial-section electron microscopy \cite{dichio2023phd,hassan2015,witvliet2021}. 
\smallskip

\paragraph{Theoretical framework.} 
% Terminology
Let us begin by clarifying the terminology. (a) Exploration refers to the act of stochastically searching the configuration space. (b) Exploitation refers to the harnessing the discovered configurations to optimise the system function. The resulting optimisation problem is defined once we specify (b.i) how the optimal states are encoded and (b.ii) how the system approaches them.

% Object
Formally, let us consider a biological system represented as a simple graph (or network) $G\in\mathcal{G}$ over $N$ nodes, unweighted, undirected, with no self loops. It can be identified with a finite, binary, symmetric and with zero-diagonal adjacency matrix  $G = \{a_{ij}\}$, where $a_{ij}\in \{0,1\}$ indicates the absence or presence of an edge within the pair of nodes, or dyad, $(ij)$. There are $L = N(N-1)/2$ dyads, hence $L$ possible edges. 
Let $P(G,t)$ be the probability of the graph $G$ at time $t$.

(a) \emph{Exploration}. Each dyad mutates its state in the time interval $\Delta t$ with rate $\mu\ge0$. A simple exploration scheme is to randomly create or dissolve edges, e.g., an edge is added if none existed or removed if present. The effect on the graph distribution is
\begin{equation}\label{e-mut}
\begin{split}
    P(G,t+&\Delta t) =\\& P(G,t) + \Delta t\mu\sum_{i<j} [P(M_{ij}G,t)-P(G,t)]  \ ,
\end{split}
\end{equation}
where $M_{ij}$ is the operator that mutates the dyad $a_{ij}$ of the graph $G$. The exploration rate $\mu$ is here constant and uniform across dyads. 

(b) \emph{Exploitation}. A functional metric $F(G):\mathcal{G}\rightarrow\mathbb{R}$ serves the purpose of representing the concept of biological function, with optimal states defined as maxima of $F$ (b.i). In the time interval $\Delta t$, we formally define exploitation as follows:
\begin{equation}\label{e-sel}
    P(G,t+\Delta t) = \frac{e^{\Delta t \varphi F(G)}}{\av{e^{\Delta t \varphi F}}_t}P(G,t)\ ,
\end{equation}
where $\av{\cdot}_t$ stands for the ensemble average at time $t$ i.e. $\av{e^{\Delta t \varphi F}}_t = \sum_{G} e^{\Delta t \varphi F(G)}P(G,t)$ and $\varphi\ge0$, the exploitation rate, is an overall scaling. Therefore, the way in which the dynamics approach the most functional (highest $F$) configurations is by exponentially increasing the probability of those graphs that have higher $F$ values than the ensemble average at time $t$ (b.ii).

We will refer to the ratio $\rho = \varphi/\mu$ as the functional pressure: $\rho\sim0$ implies a dynamic dominated by randomness, similar to a random walk in the graph space $\mathcal{G}$, while $\rho\to\infty$ corresponds to the limit of perfect exploitation, where only the most functional graph configurations have non-negligible probabilities. 

In SM-II we introduce four simple models, namely the cases of no exploitation, edge penalty, edge covariate and distance-like $F$ metric. We show that these cases can be treated analytically and offer a formal understanding of the intuitions (i) that the optimal states implied by a $F$ metric are not strictly attainable as long as $\mu\ne0$, and (ii) that the functional pressure $\rho$ controls not only the rate of approach to the maxima of $F$ but also the final stationary state.

The above framework closely mimics a Darwinian evolution driven by mutations and fitness-based natural selection, in the infinite population limit \cite{neher2011,dichio2023}.
Eq.(\ref{e-mut}-\ref{e-sel}) can be regarded as an algorithm, inspired by evolution, that (a) uses random choices to (b) direct an exploitative search for solving an optimisation problem. In this sense, it is similar to a genetic algorithm \cite{goldberg1989}, see SM-I. We further explore the parallel with the evolutionary dynamics to design simulations based on eq.(\ref{e-mut}-\ref{e-sel}), concisely described in SM-III.

$F$ is shaped by the environment and has the role of mapping the functional requirements of the biological system onto the configuration space. As a consequence, the particular form of $F$ depends on the specific context and system. In general, there are no requirements on the properties of $F$, which could be regarded as a black box which returns a real number for each possible input (graph). In this work, however, we will study a white-box $F$ metric which admits a mathematical formulation. In particular, we will describe the state of a graph $G$ by a set of \emph{sufficient statistics} $\bm{x}(G)\in\mathbb{R}^r$ and in this latter space the $F:\mathbb{R}^r\to\mathbb{R}$ will be formally defined.
\smallskip

\paragraph{C. elegans brain maturation.}
% General
In the following, we will consider the so-called brain wiring problem \cite{hassan2015}, i.e., how the structural complexity of a natural brain arises during the development of an organism. Answering the brain wiring problem is an open challenge in neuroscience and entails tackling at least two kind of questions: (i) what drives the brain maturation (\emph{structural principles}) (ii) which is the driving algorithm (\emph{dynamical principles}). Here, we will formulate them in terms of a $F$-metric and EE dynamics, respectively. In particular, the latter is consistent with three essential and general features of the brain wiring dynamics, which are
\begin{itemize}
    \item[(a)] \emph{functionally robust} -- the adult brains are capable of supporting the functions that sustain the life of an organism;
    \item[(b)] \emph{not hardwired} -- genetically encoded developmental algorithms give rise to similar yet non-identical structures, resulting in high inter-individual variability \cite{hassan2015};
    \item[(c)] \emph{self-referential} -- the updating rules evolve in time, as a function of the state and therefore of the history of the system \cite{goldenfeld2011}.
\end{itemize}

% Data
To tackle the brain wiring problem, a natural choice is to consider that of the nematode \ce\ \cite{white1986}, SM-V.A. This is the only organism for which a comprehensive map of neuronal connections within a brain has been reconstructed across development \cite{witvliet2021}. The dataset consists of $8$ fully reconstructed brains of the hermaphrodite \ce, obtained from different isogenic individuals at different developmental ages, including one at birth and two adults ($t\sim\SI{45}{\hour}$ after birth), SM-V.B. We consider the unweighted and undirected networks of chemical synapses between sensory, inter, motor and modulatory neurons ($161-180$ nodes, $617-1669$ edges). This choice of representation is motivated by the statistical properties of the adult \ce\ connectome \cite{varshney2011,cook2019}, along with the effort to devise a simplified growth model, a critical discussion can be found in SM-V.B.

\begin{figure}
    \centering
    \includegraphics[width=\columnwidth]{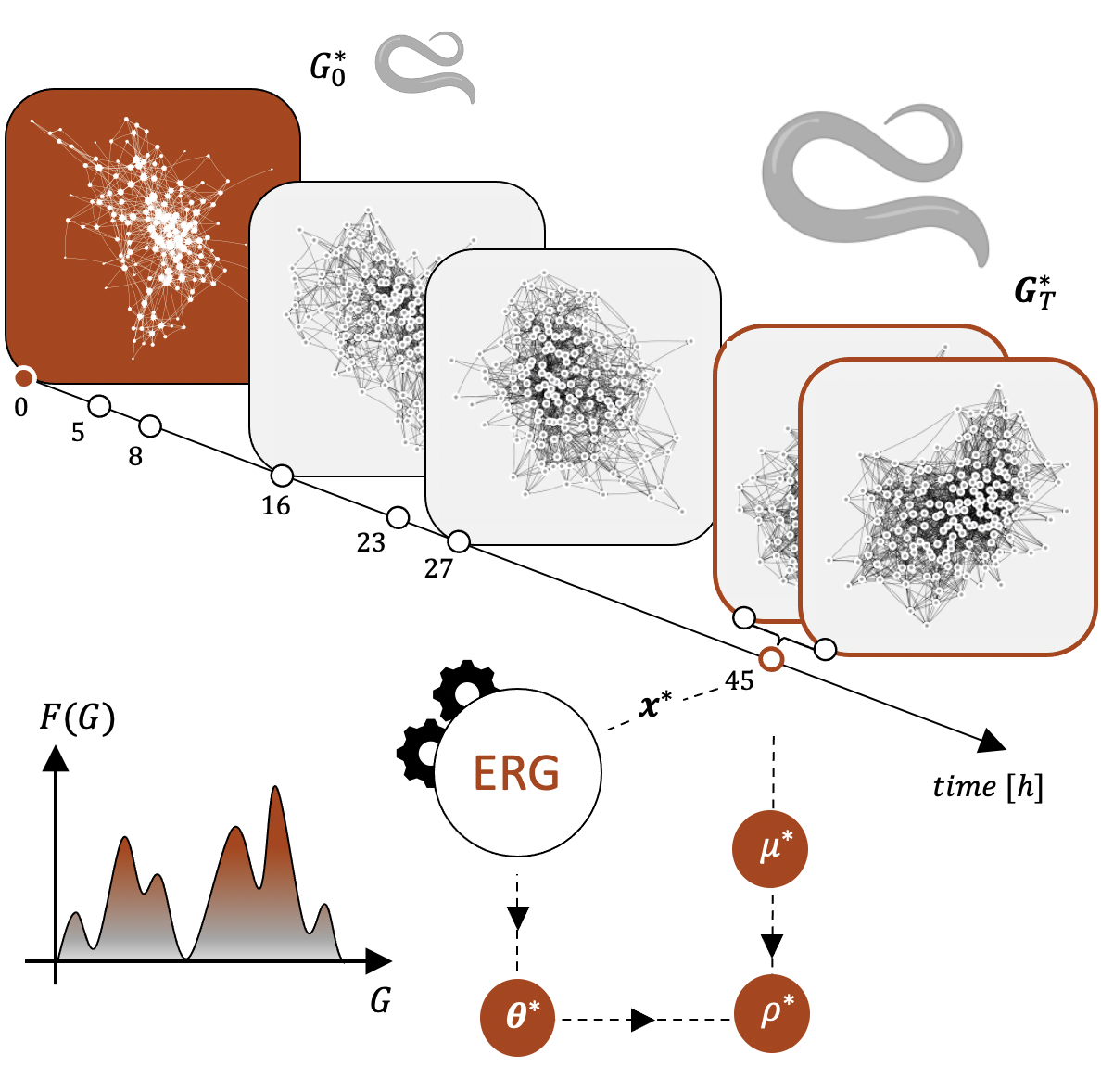}
    \caption{EE dynamics for the \ce\ brain maturation. We consider the eight snapshots of the worm network of chemical synapses, at different developmental ages $t=\SIlist{5;8;16;23;27;45}{\hour}$ (two adults), from \cite{witvliet2021} -- some are omitted for visual clarity. The birth configuration is fixed as the starting point of the dynamics. The two adult snapshots are used to infer (i) the topography of the functional landscape $F(G)$ (bottom left), encoded in the set $\bm{\theta}^*$ -- ERG inference starting from the empirical statistics $\bm{x}^*\equiv\bm{x}(\bm{G}_T^*)$; (ii) the EE parameters, i.e., the exploration rate $\mu^*$ and the functional pressure $\rho^*$, see SM-VI.
    }
    \label{f-overview}
\end{figure}

% Motivation / Interpretation
The developmental principles that guide the \emph{C.elegans} brain maturation are not entirely known. On the one hand, approximately $43\%$ of the synaptic connections between neurons are not conserved among genetically identical individuals, suggesting a prominent role of stochasticity in the brain wiring \cite{witvliet2021,hassan2015}. Conversely, the diverse range of behaviors exhibited by adult \emph{C.elegans} \cite{bono2005} demands functional selection. Our EE framework captures these two tendencies simultaneously. An overview of the approach is illustrated in fig.(\ref{f-overview}). A micro-level interpretation of eq.(\ref{e-mut}-\ref{e-sel}) for the wiring dynamics of the individual neurons is extensively discussed in SM-V.C. % RL B.1

% Methods - Choice of statistics
A preliminary step of our modelling approach is the characterization of the worm brain by a set of sufficient statistics $\bm{x}(G)$. Based on recent evidence \cite{pathak2020,azulay2016}, we consider a parsimonious representation in which:
\begin{equation}\label{e-stats}
  \bm{x}(G) = 
  \left[
    \begin{array}{c}
    \sum_{k>0} w^{(k)}_{\tau_{d}}\ x_{d}^{(k)}(G) \\[0.5em]
    \sum_{k>0} w^{(k)}_{\tau_{esp}}\ x_{esp}^{(k)}(G) 
    \end{array}
  \right] \ ,
\end{equation}
where $x_d^{(k)}$ and $x_{esp}^{(k)}$ are the number of nodes with degree $k$ and the number of connected dyads sharing exactly $k$ partners, respectively. The coefficients are $w^{(k)}_{\alpha} = e^{\alpha}\{1-\big(1-e^{-\alpha}\big)^{k}\}$, with $\alpha=\tau_d,\tau_{esp} >0$ decay parameters. In other words, these statistics are linear combinations of the degree and edgewise shared partner distributions. They yield a model that is both realistic and computationally tractable \cite{dichio2022}, SM-IV.  
More specifically, the first statistic is called geometrically weighted degree (\texttt{gwd}) and encodes the information, e.g., on the presence/absence of hub-nodes in the graph, well documented in the case of the \emph{C.elegans} \cite{pathak2020,witvliet2021}. The second statistic is called geometrically weighted edgewise shared partner (\texttt{gwesp}) and is a proxy for a triadic-closure phenomenon in the graph, i.e., pairs of nodes that have links to one or more common neighbours have a higher chance of being connected to each other. The latter could in turn result from a tendency of the network to segregate into densely connected modules.  \cite{witvliet2021,pathak2020}.

% Methods - ERGM inference
Given a choice of statistics as in eq.(\ref{e-stats}), we can characterize any observed graph $G^*$ within the inferential framework of exponential random graph (ERG) models \cite{cimini2019,dichio2022}, SM-IV. Accordingly, the ensemble $\mathcal{G}$ is endowed with a maximum entropy probability distribution
\begin{equation}\label{e-ERGM-MS}
    P_{ERGM}(G|\bm{\theta}) = e^{-\mathcal{H}(G,\bm{\theta})} / \sum_{\Tilde{G}\in\mathcal{G}}e^{-\mathcal{H}(\Tilde{G},\bm{\theta})}
\end{equation}
where $\mathcal{H}(G,\bm{\theta}) = -\bm{\theta}\cdot\bm{x}(G)$ is the Hamiltonian. Given an observed graph $G^*$, the vector of parameters $\bm{\theta}^*$ can be inferred as approximate solution the maximum likelihood estimation problem $\bm{\theta}^* = \argmax_{\bm{\theta}} \ \log P(G^*|\bm{\theta}).$ The inferred parameters quantify the contribution of the associated statistics to the structure of the observed graph. 
For example, if the statistic $x_{\alpha}$ is non-negative, positive $\theta_{\alpha}^*$ imply the existence of a bias towards graphs with higher-than-random values of $x_{\alpha}$ -- given the rest of the model $\sum_{\beta\ne\alpha}\theta_{\beta}^*x_{\beta}$ \cite{dichio2022}, SM-IV.

% Methods - F-metric
We are now in the position to propose the following $F$ metric for the \ce\ brain maturation 
\begin{equation}\label{e-F-ce}
    F(G) = \bm{\theta}^*\cdot\bm{x}(G)\ ,
\end{equation}
where $\bm{x}(G)$ are defined in eq.(\ref{e-stats}). The parameters $\bm{\theta}^*$ are obtained from the ERG inference in the adult stage, so that the correct (functional) balance of model statistics can be achieved at the end of the developmental process. The topography of the \emph{functional landscape} is genetically encoded and results from the combined effect of physical, genetic and functional constraints. 

In particular, we consider the average estimated parameters from the two adult worms $\bm{G}_{T}^*=(G_{T,1}^{*},G_{T,2}^{*})$ and obtain $\theta^*_{gwd} = 0.44$, $\theta_{gwesp}^* = 0.58$, SM-VI.A. EE dynamics based on eq.(\ref{e-F-ce}) will favor both the emergence of hubs and of a triadic closure behaviour since, by virtue of the positive values of the linear parameters, higher values of the statistics in eq.(\ref{e-stats}) will imply higher $F$ values. This is in line with experimental observations that, during development, hub neurons at birth get more inputs and that the overall modularity of the \ce brain network increases \cite{witvliet2021}.

We can now proceed to model the developmental dynamics by setting appropriate boundary conditions, and the EE parameters of the dynamics: the exploration rate and the functional pressure. As the argument goes, $\mu$ and $\rho$ are characteristic of the specific instance (\ce) of the biological process (brain wiring), they are genetically encoded and therefore result from the evolutionary history of the species.

% Methods - Boundary conditions
We set the graph $G_{0}^*$ corresponding to the network at birth as the starting point of the dynamics, $P(G=G_{0}^*,0)=1$. In fact, (i) as reported in \cite{witvliet2021}, the brain morphology at birth serves as the structural foundation upon which the adult connectivity unfolds. Moreover, (ii) an implicit assumption of the EE graph dynamics is the \emph{functional homogeneity}, i.e., that the same $F$ metric holds true throughout the whole dynamics. This assumption is likely to be violated before hatching (birth), during the embryonic stage, where a different growth regime of the nervous system has been observed \cite{nicosia2013}.

\begin{figure*}
\includegraphics[width=\linewidth]{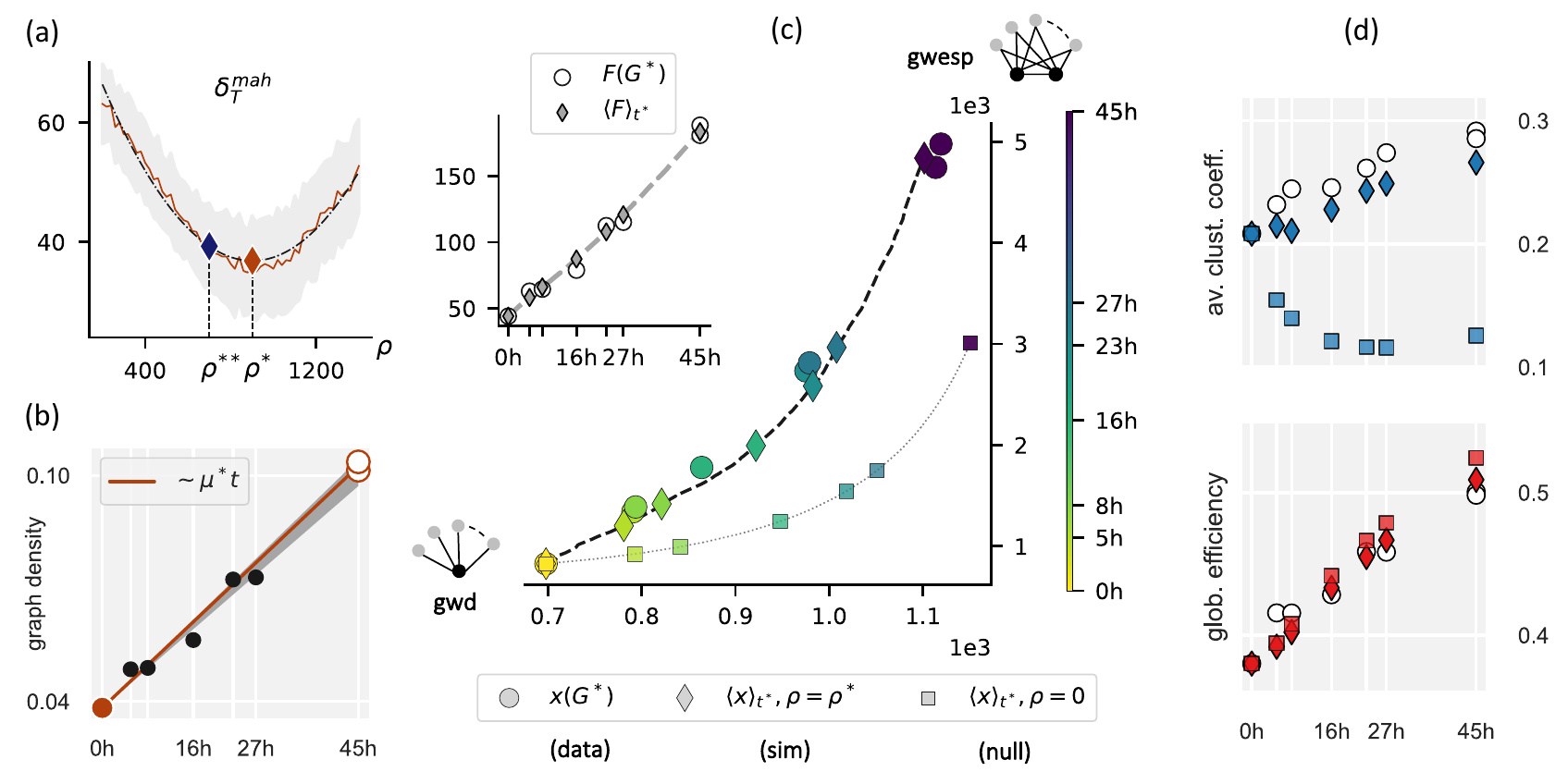}
\caption{\label{f-results} Tracking down the \ce\ brain maturation. (a) We run $100$ simulations $\forall \rho\in \{200+20\ i, 0\le i\le60 \}$. For each $\rho$, we compute the mean and standard deviation of $\delta_T^{mah}$ (red line and shaded area, respectively). We fit the data with a quadratic curve (dash-dotted line) and take the abscissa $\rho^*$ of its minimum (red diamond) as estimation of the functional pressure. On the same curve, we show the value (blue diamond) corresponding to the abscissa $\rho^{**}$ we get by minimizing the sum of the Mahalanobis distances over all experimental time-points, SM-VI.B. The two overlap within the error bars. (b) The exploration rate is calculated by using the first and last time point (average). The shaded area corresponds to the estimation by a linear fit over the whole time series. (c) One simulation run with $\mu^*,\rho^*$. Main: the trajectory in the space of statistics (\texttt{gwd, gwesp}). Experimental data (circles) are closely tracked by our simulations (dashed line, diamonds highlighting the \emph{observed} time points $\bm{t^*}$). The trajectory of a null model with $\rho=0$ is also shown (dotted line, squares)  Inset: The simulated and experimental time course of $F$ in time. (d) Feature generalisation. The temporal trajectory of the average clustering coefficient and global efficiency, markers as described in (c). See also SM-VI.C.}
\end{figure*}

% Methods + Results - Exploration rate
Throughout development, the removal of synaptic connections happens rarely \cite{witvliet2021}. Accordingly, we modify the mutation scheme described in eq.(\ref{e-mut}) by restricting the removal of edges. By Occam's razor, we assume a constant exploration rate
\begin{equation}\label{e-mut-rate}
    \mu^*=\frac{1}{TL}\sum_{i<j}\big[\bar{a}_{ij}(\bm{G}_{T}^*)-a_{ij}(G_{0}^*)\big] = 1.43\times10^{-3}  h^{-1} ,
\end{equation}
where $T=45h$ is the adult age, $L=N(N-1)/2$ is the number of dyads of the adult brain graphs ($N=180$), $\sum_{i<j}\bar{a}_{ij}(\bm{G}_{T}^*)$ is the average number of edges between the two adult worms and $\sum_{i<j}a_{ij}(G_{0}^*)$ is the number of edges at birth. At the end of the developmental dynamics, $\mu^*$ is assumed to drop to zero.

% Methods + Results - Functional pressure
We are left with only one free parameter, i.e., the functional pressure $\rho=\varphi/\mu>0$. 
Nonzero $\rho$ describe the scenario in which new connections emerge primarily in locations where they result in an enhanced system function. Its value must be biologically regulated to ensure the development of adequately specialized functional circuits prior to reaching the adult stage. Therefore, we use the corresponding degree of freedom to inform the EE graph dynamics about the age of the adulthood. In particular, we set $\rho^* = \min_{\rho}\delta^{mah}_T \ ,$ where the quantity to be minimized is the Mahalanobis distance \cite{mahalanobis2018}, at time $T$, between the two-dimensional ensemble distribution of the graph statistics and the average experimental values, SM-VI.B. 

% Result - Mah convex function
In fig.(\ref{f-results}a) we show that that $\delta^{mah}_T$ is a convex function of the functional pressure $\rho$. Both insufficient and excessive $\rho$ lead to the ensemble distribution diverging from the experimental values. The minimization procedure yields $\rho^* = 9.017\times10^2$ ($R^2=.98$). 

% Result - T45 predicts time-series
Notably, in estimating the dynamical parameters we have relied solely on the \ce\ brain graphs at birth and in the adult stage. We can ask if and how the estimation would change when considering the whole available data, which include the developmental time-points at $\SIlist{5;8;16;23;27;45}{\hour}$ after birth. A linear fit of the growth of the number of edges based on the whole time series yields $\mu^{**}=(1.39\pm0.08)\times10^{-3}h^{-1}$. This estimation is compatible with the value $\mu^*$ in eq.(\ref{e-mut-rate}), fig.(\ref{f-results}b). As for the functional pressure, we can define an equivalent minimization problem where the Mahalanobis distance is summed over all experimental time points, yielding a value $\rho^{**}$ that is statistically consistent with $\rho^*$, fig.(\ref{f-results}a), SM-VI.B. 

This hints that (i) given the model eq.(\ref{e-F-ce}), the assumption of functional homogeneity for the worm brain wiring dynamics holds true and (ii) the EE graph dynamics, informed about birth and adulthood, capture the entire developmental trajectory.

% Results - one sim
To further investigate this result, we can fix $\mu^*,\rho^*$ and look at individual simulations of the brain growth, as in fig.(\ref{f-results}c). For comparison, we also plot the results of a null model where $\rho=0$, i.e., a random graph growth with exploration rate $\mu^*$. As could be expected, the adult stage is correctly reached in terms of the model statistics eq.(\ref{e-stats}) and, by consequence, of the $F$-metric. Notably, however, our simulations approximate \emph{en passant} the other observed developmental ages, which we have used nowhere in inferring the parameters. This opens up the possibility of using our framework to reconstruct also those stages of brain maturation for which no data are available. 

% Results - Gof  + being humble
The model described so far is a simple, low-dimensional model of the underlying biological dynamics. Almost by construction, a choice as simple as eq.(\ref{e-stats}) is unlikely to capture the finer-scale topological details of the observed graphs. Analogously, an exploration dynamics as simple as eq.(\ref{e-mut-rate}) cannot capture transient dynamic patterns. Yet, we can meaningfully ask to what extent the EE graph dynamics based on the features eq.(\ref{e-stats}) reproduces other network properties not included in the model formulation, SM-VI.C. In fig.(\ref{f-results}d) we show that our model retrieves the  propensity of the \ce\ brain networks to exhibit relatively high efficiency (like random graphs) and clustering (unlike random graphs) \cite{latora2017}.

\paragraph{Discussion.}
To summarize, we have presented a parsimonious, interpretable framework for the dynamics of networked biological systems. It is built upon the dynamical principle of the exploration-exploitation (EE) paradigm, which is general. 
It serves as theoretical scaffolding for formulating specific dynamical models, which must be tailored to the biological system. We have used it here to model the growth of the \ce\ connectome, from birth to adulthood. Notably, our results suggest that the knowledge of the birth and adult age is sufficient for the EE graph dynamics to describe the whole developmental trajectory. We speculate that the same may be true for the connectomes of other living systems \cite{hildebrand2017,scheffer2020,abbott2020}, for which no such data as the developmental trajectory are available to date. This hypothesis is poised for experimental validation in the near future.

Our model should be regarded as a first step towards a more detailed understanding of the brain maturation. To this end, the framework here presented supports straightforward extensions to more complex exploration schemes, accounting for non-uniform synapse addition, directed flow of synaptic information, neuron-specific information, homophily effects, and physical or functional constraints \cite{witvliet2021,pathak2020}. A detailed discussion of the possible model extensions can be found in SM-VII.A. Beyond structural connectivity, it would be interesting to study under the same lens the \ce\ brain functional connectivity, recently mapped for the adult stage \cite{randi2022}, where there exists a closer correlation between the notion of biological function and the topology of the graph. 

Zooming out, our framework can be broadly used to study the dynamics of complex systems arising from the interplay between (i) the variability fueled by a stochastic search of the configuration space and (ii) the state-dependent optimisation of an objective function -- we propose three examples in SM-VII.B. Importantly, as showcased here, this can be done by introducing only a very limited number of interpretable parameters \cite{dyson2004}\footnote{Additional discussion can be found in Supplemental Material at [url], which includes Refs. \cite{hamilton2021,manrubia2021,neher2011,dichio2023,goldberg1989,zanini2012,mauri2021,albert2002,jaynes1957,cimini2019,dichio2022,geyer1992,krivitsky2022,schweinberger2020,hunter2007,bentley2016,cook2019,white1986,skuhersky2022,pathak2020,rapti2020,witvliet2021,varshney2011,latora2017,colon2009,mahalanobis2018,kessy2018,hildebrand2017,dorkenwald2023,lee2016,siegel2018,dancause2005,mislove2007,reid2012,dichio2023git}.}.
\medskip

\begin{acknowledgments}
We thank E Aurell, E Mauri, D Battaglia and M Josserand for the many useful discussions. FDVF acknowledges support from the European Research Council (ERC), Grant Agreement No. 864729.
\end{acknowledgments}

\bibliography{mybib}

\end{document}

%\preprint{AIP/123-QED}

\title[Supplemental Material]{Supplemental Material \\ The exploration-exploitation paradigm for networked biological systems}% Force line breaks with \\
%\thanks{}

\author{Vito Dichio}
\author{Fabrizio De Vico Fallani}%
\affiliation{ 
Sorbonne Universite, Paris Brain Institute - ICM, CNRS, Inria, Inserm, AP-HP, Hopital de la Pitie Salpêtriere, F-75013, Paris, France\looseness=-1}

\date{\today}

\maketitle
\tableofcontents
\clearpage

In this document, we present additional information to supplement the discussion in the main manuscript, which we will refer to as \ms. All references (equations, figures, tables, listings) in the Supplemental Material are labeled with roman numbers. 

\section{Darwin and the others}\label{s-darwin}

The theoretical framework presented in \ms\ is constructed by abstracting the core algorithm of Darwinian evolution to a general exploration-exploitation (EE) dynamic. To elucidate this analogy, we here provide a glimpse of the main features of the former.

% Darwinian evolution
Darwinian evolution, or simply Darwinism, is the widely-accepted theory of biological evolution introduced by the English naturalist Charles Darwin in his seminal work, \emph{On the Origin of Species} (1859). The problem considered is that of a population (a group of organisms of a species) in the same environment, that reproduce across successive generations. The key ingredient of evolution is \emph{inheritance}: offspring inherit traits (phenotype) from their parents through genetic information passed down from generation to generation. The different forms in which each gene may exist are called alleles, the pool of alleles of an organism is its genotype. At the population level, two opposing principles cooperate in defining the evolutionary dynamics. 
\begin{enumerate}[(a)]
    \item Genetic \emph{variation}. Stochastic events drive the generation of genetic diversity within a population by introducing variability in the genetic makeup of individuals. This diversity arises primarily through two mechanisms: mutations, which involve random alterations in the genotype, and recombinations, which involve the exchange and shuffling of genetic material between homologous genotypes.
    \item Natural \emph{selection}. It acts upon this genetic variation by favoring individuals with traits that confer a reproductive advantage in a given environment. As a consequence of selection, populations gradually become better adapted to the environment (adaptation).
\end{enumerate}

% The NS model
General, up-to-date discussions of these concepts can be found in \cite{hamilton2021,manrubia2021}. Among the existing theoretical approaches to evolutionary dynamics, we mention in particular the one by \emph{Neher \& Shraiman}, i.e., the framework of \emph{statistical genetics}, proposed in \cite{neher2011} and recently reviewed in \cite{dichio2023}. Here, genotypes are modelled as Ising spin-chains $g = \{\sigma_1,\dots,\sigma_L\}$, where $\sigma_i\in\{-1,1\}$ and $L$ is the fixed genome length. The evolutionary dynamic is framed in terms of a master equation for $P(g)$, the probability distribution in the genotype space, which changes under the effect of mutations (spin swaps), recombinations (reshuffling of a pair of genomic chains) and selection. The latter is based on a fitness function $F(g)\in\mathbb{R}$, which quantifies the aptness of an individual to the environment: at any time, those individuals that are more apt to the environment (higher $F$) than the others in the population will have a higher chance to pass their genetic information to the next generation. Note that this implies that selection is state-dependent, since what matters is not the aptness of an individual the environment \emph{per se} but with respect to that of all others in the population.

% Darwin's algorithm
In formulating the exploration-exploitation (EE) dynamic, we borrowed the formal structure of the approach above described. The key logical step is to recognize that the evolutionary dynamic is a particular instance of a general EE dynamic in which (i) the configuration space is that of all possible genotypes (ii) exploration is realized by genetic mutations and recombinations and (iii) the exploitation is driven by natural selection. If the interpretations we attach to the concepts of exploration and exploitation are context-dependent, the algorithm coded by evolution to solve its exploration-exploitation problem can be adapted and used elsewhere.

% Genetic algorithms
This same line of ideas underlies the family of \emph{genetic algorithms}, a broad class of computational methods, designed in a way to mimic the evolutionary process to solve complex optimisation problems \cite{goldberg1989}. An evolutionary process is replicated \emph{in silico} in order to optimise a target objective function. To this end, $M$ individuals (instances of the system) of a population are iteratively mutated, recombined and selected, until the appropriate termination criterion is met. 
Albeit powerful, this approach is inherently algorithmic, meaning that its interpretation as a direct representation of a real-world process (other than the evolutionary dynamic) is problematic. In particular, outside the context of evolutionary (population) dynamics, it is unclear what the recombinations should correspond to.

% Our EE dynamic
In order to retain the potential for interpretation, in eq.(\ref{e-mut}-\ref{e-sel}) we implement an EE dynamic where exploration is solely achieved by random mutations, while no recombination-like mechanism exists. 
In other words, when borrowing from the Darwin's algorithm, we limit ourselves to those mechanisms that are not specific of the evolutionary case \cite{dichio2023phd}. Both mutations and fitness-based selection indeed offer a straightforward abstraction - a dictionary can be found in tab.(\ref{t-EE-evol}) - and an interpretation in different contexts. In sec. \ref{ss-interpr} the case of the \ce\ brain maturation is discussed.   
\begin{table}[htbp]
  \centering
  \caption{Translation dictionary between exploration-exploitation EE graph dynamic (left) and evolutionary dynamic (right). We group by colour those terms that refer to the structure of the configuration and state spaces (top), to the the structure of the dynamics (middle) and to the dynamic parameters (bottom). No equivalent mechanism to genetic recombinations exists in the EE graph dynamic. *See also sec. \ref{s-sim}.}
  \begin{tabularx}{.62\textwidth}{>{\raggedright\arraybackslash\hsize=1.3\hsize}X>{\centering\arraybackslash\hsize=0.4\hsize}X>{\raggedleft\arraybackslash\hsize=1.3\hsize}X}
    \textbf{EE graph dyn.} & \textbf{Notation} & \textbf{Evolutionary dyn.} \\
    \toprule
    \rowcolor[HTML]{DFD7BF}
    graph space  & $\mathcal{G}$ & genotype space \\
    \rowcolor[HTML]{DFD7BF}
    number of dyads  & $L$ & genome length \\
    \rowcolor[HTML]{DFD7BF}
    graph (unweighted, undirected) & $G/g$ & genotype (biallelic)\\
    \rowcolor[HTML]{DFD7BF}
    graph statistics & $\bm{x}$ & phenotype (traits)\\
    \rowcolor[HTML]{DFD7BF}
    dyad (bit-like)  & $a_{ij}/\sigma_{ij}$ & locus (spin-like) \\ \addlinespace[3pt]
    \rowcolor[HTML]{F2EAD3}
    time window  & $T$ & \# generation \\
    \rowcolor[HTML]{F2EAD3}
    \# samples$^*$  & $M$ & population size \\
    \rowcolor[HTML]{F2EAD3}
    biological function ($F$ metric)  & $F$ & fitness function  \\
    \addlinespace[3pt]
    \rowcolor[HTML]{F5F5F5}
    exploitation r.  & $\varphi$ & natural selection r. \\
    \rowcolor[HTML]{F5F5F5}
    exploration r.  & $\mu$ & mutation r. \\
    \rowcolor[HTML]{F5F5F5}
    $\times$  & $r$ & recombination r.\\ 
    \bottomrule
  \end{tabularx}
  \label{t-EE-evol}
\end{table}
\section{The math of simple scenarios}
Let us start by combining the EE graph dynamic eq.(\ref{e-mut}-\ref{e-sel}) in a single formula:

\begin{equation}\label{e-me}
P(G,t+\Delta t) = P(G,t) + \Delta t\mu\sum_{i<j} [P(M_{ij}G,t)-P(G,t)] + \Bigg[\frac{e^{\Delta t \varphi F(G)}}{\av{e^{\Delta t \varphi F}}_t}-1\Bigg]P(G,t) \ . 
\end{equation}
where all symbols hold as defined in \ms. In order to simplify the calculations, it is convenient to define the \emph{spin} variables $\sigma_{ij}={\pm1}$ which are related to the dyadic variables $a_{ij}$ in \ms by the relations: 
\begin{equation}\label{e-spdy}
    \sigma_{ij} = 2a_{ij}-1\ , \quad \quad a_{ij} = \frac{1+\sigma_{ij}}{2} \ .
\end{equation}
It is important to note that in the context of graphs, the dyads, not the nodes, assume the role analogous to spins in classical statistical mechanics. In a similar fashion, we can define the \emph{average magnetization} $m\in[-1,1]$ as the spin analogous of the \emph{average graph density} $d\in[0,1]$ as 

\begin{equation}\label{e-magden}
    m = \frac{1}{L}\sum_{i<j}\av{\sigma_{ij}} = 2d-1\ ,\quad \quad d = \frac{1}{L} \sum_{i<j} \av{a_{ij}} = \frac{1+m}{2}\ .
\end{equation}

The action of an operator $M_{ij}$ that mutates the state of the dyad $a_{ij}$ from $1\rightarrow0$ or $0\rightarrow1$ (edge toggle) can compactly be expressed as $M_{ij}: \sigma_{ij}\rightarrow-\sigma_{ij}$, a spin flip. We will consider in this section the regime of weak exploitation in which $\Delta t\varphi F(G) \ll 1$. A continuous-time description of eq.(\ref{e-me}) can then be given:
\begin{equation}\label{e-mec}
    \frac{d}{dt}P(G,t) = \mu\sum_{i<j} [P(M_{ij}G,t)-P(G,t)] + \varphi [F(G)-\av{F}_t]P(G,t) \ ,
\end{equation}
where we have used $e^{\pm x}\sim 1\pm x$ for $x\ll1$. Eq.(\ref{e-mec}) allows to compute the dynamic of the expected value of any graph-observable $O(G):\mathcal{G}\rightarrow\mathbb{R}$. In fact:
\begin{equation}
    \frac{d}{dt}\av{O}_t = \frac{d}{dt}\sum_{G}O(G)P(G,t) = \sum_{G}O(G)\ \boxed{\frac{d}{dt}P(G,t) } \ ,
\end{equation}
where we have introduced the shorthand $\sum_{G} = \sum_{\sigma_{11}=\pm1}\sum_{\sigma_{12}=\pm1}\dots\sum_{\sigma_{LL}=\pm1}$. Let us now examine some simple cases that are analytically tractable.

\subsection{No exploitation}
In the trivial case in which $F=const$ (no exploitation), the dynamic of the system is a random walk in the graph space, whose speed is tuned by the rate $\mu$. We can easily compute the expected value of any spin variable as:
\begin{equation}\label{e-EE-no-exp}
    \frac{d}{dt}\av{\sigma_{ij}}_t \stackrel{}{=} \sum_{G} \sigma_{ij}\ \mu\sum_{k<l} [P(M_{kl}G,t)-P(G,t)] = \mu\Big[ \sum_G\sigma_{ij}P(M_{ij}G,t)- \sum_G\sigma_{ij}P(G,t)\Big]\stackrel{}{=}  -2\mu\av{\sigma_{ij}}_t\ ,
\end{equation}
where in the last step we have used $\sum_G\sigma_{ij}P(M_{ij}G,t) = \sum_{G} -\sigma_{ij}P(G,t)$. Eq.(\ref{e-EE-no-exp}) is the differential equation of an exponential decay with characteristic time $(2\mu)^{-1}$, yielding the solution $\av{\sigma_{ij}}_t=e^{-2\mu t}\av{\sigma_{ij}}_{t_0}$. In words, after a sufficiently long time, any memory of the initial condition is lost. Each local spin variable converges to a state that is statistically analogous to the outcome of a fair coin toss. Since the dynamics of each variable $\sigma_{ij}$ are independent, the magnetization follows the same exponential decay $m_t=e^{-2\mu t}m_{t_0}$. In terms of graph density, this implies that starting from any initial condition, the average state of the system melts down in an Erd\H{o}s-Rényi random graph where each edge exists with probability $p=1/2$.

\subsection{Edge penalty}\label{s-epen}
\begin{figure}
\includegraphics[width=.6\columnwidth]{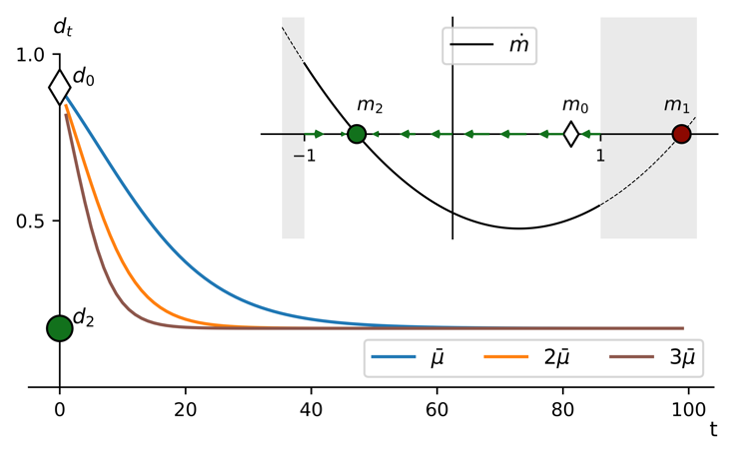}
\caption{Analytical solution of the exploration-exploitation dynamic based on eq.(\ref{e-fep-d}), decoupling approximation. Here $N=10$ ($L=45$), $\bar{\mu}=1/L, \bar{\rho}=2\times10^2$, $d_0=0.9$. Inset: plot of $\dot{m}=f(m)$, where $f$ is specified by eq.(\ref{e-fep-d-approx}). The dynamic has two fixed points, corresponding to the solutions of $f(m)=0$, labelled as $m_1$ (red, unstable) and $m_2$ (green, stable). Shaded area corresponds to the inaccessible regions for $m\in[-1,1]$. The initial condition $m_0$ (white diamond) and the one-dimensional vector field (green arrows) are indicated. Main: dynamic of the average graph density $d_t$ as obtained from eq.(\ref{e-fep-sol}) using eq.(\ref{e-magden}). The value of the asymptotic state $d_2$ solely depends on $L, \bar{\rho}$ (held fixed) while the rapidity of the approach to it additionally depends on the exploration rate $\mu$.
}\label{f-ed}
\end{figure}

A simple exploitation scheme is the one in which the only graph sufficient statistic is the number of edges and each existing edge results in a fixed penalty, i.e.,
\begin{equation}\label{e-fep}
    F(G) = -\frac{1}{L}\sum_{i<j}a_{ij}\ ,
\end{equation}
where $L=N(N-1)/2$. The case of a fixed benefit per edge is equivalent to eq.(\ref{e-fep}), modulo a minus sign. Using eq.(\ref{e-spdy}), we can express the exploitation component of eq.(\ref{e-mec}) in term of spin variables as
\begin{equation}
    [F(G)-\av{F}_t]P(G,t) = -\frac{1}{2L}\Big[\sum_{i<j}(\sigma_{ij}-\av{\sigma_{ij}}_t)\Big]P(G,t)\ ,
\end{equation}
and again compute the dynamic of the expected value for any spin variable as
\begin{equation}\label{e-fep-d}
    \frac{d}{dt}\av{\sigma_{ij}}_t = -2\mu\av{\sigma_{ij}}_t - \frac{\varphi}{2L}\sum_{k<l} \big[\av{\sigma_{ij}\sigma_{kl}}_t-\av{\sigma_{ij}}_t\av{\sigma_{kl}}_t\big]
\end{equation}
Differently from the previous case, the dynamics of the average spin variables are now coupled by the sum of covariances $C_{ij;kl}=\av{\sigma_{ij}\sigma_{kl}}-\av{\sigma_{ij}}\av{\sigma_{kl}}$ in the last term. The calculation simplifies if we restrict to the case in which the spin-covariance matrix has approximately a diagonal form, i.e., $C_{ij;ij}=1-\av{\sigma_{ij}}^2\sim\mathcal{O}(1)$ and $C_{ij;kl}\sim\mathcal{O}(\epsilon)$ for $ij\ne kl$ (\emph{decoupling approximation}). Eq.(\ref{e-fep-d}) becomes
\begin{equation}\label{e-fep-d-approx}
    \frac{d}{dt}\av{\sigma_{ij}}_t \sim -2\mu\av{\sigma_{ij}}_t - \frac{\varphi}{2L}\big[1-\av{\sigma_{ij}}^2_t\big]\ ,
\end{equation}
valid as long as $L\epsilon\ll1$, where $\epsilon$ is the order of magnitude of the off-diagonal spin covariances. In practice, this means either small graph sizes (small $L$) or mild functional pressures $\rho=\varphi/\mu$, for which the dynamic approaches that of the previous case of no exploitation (small $\epsilon$). Under the decoupling approximation the same exact equation as eq.(\ref{e-fep-d-approx}) can be written for the magnetization $m_t$ and integrated explicitly by partial fractions, yielding the solution:
\begin{equation}\label{e-fep-sol}
    m_t = m_2 \Bigg[ 1+ \frac{m_1/m_2-1}{1+\frac{m_1-m_0}{m_0-m_2}\ e^{2\mu t\sqrt{1+(2L/\rho)^{-2}}}} \Bigg]\ ,
\end{equation}
where $m_0=m_{t_0}$ and $m_{1/2}=2L/\rho\ \pm\sqrt{1+(2L/\rho)^{2}}$ are the fixed points of the dynamic. Eq.(\ref{e-fep-sol}) is an exponential relaxation dynamic to the stable fixed point $m_2$, hence $m_{t} \rightarrow m_2$ for $t\rightarrow\infty$. The magnitude of the asymptotic value $m_2$ depends exclusively on the ratio between the number of dyads $L$ and the functional pressure $\rho$. For fixed $\rho$, the rapidity of the approach to it additionally depends on the specific value of the mutation rate $\mu$, fig.(\ref{f-ed}). The edge count being the only sufficient statistic of this problem, eq.(\ref{e-fep-sol}) completely determines the dynamic. We note that for $\rho\rightarrow0$ we have $m_2\rightarrow 0$ similarly to the previous section, while for $\rho\rightarrow\infty$ we have $m_2 \rightarrow-1$, corresponding to an empty graph. For any $\rho<\infty$ the asymptotic value $m_2$ strikes a balance between the strengths of the exploration and exploitation mechanisms. 

\subsection{Edge covariate}
A straightforward generalization of the case discussed in this section is that to the case in which the $F$ is the sum of an edge covariate, i.e., 
\begin{equation}\label{e-edgecov}
    F(G) = \frac{1}{L}\sum_{i<j}a_{ij}\gamma_{ij}\ ,
\end{equation}
\noindent where $\Gamma=\{\gamma_{ij}\}$ is a $N\times N$ real-valued matrix. Each entry of $\Gamma$ contributes to the $F$ metric only when the corresponding edge exists. For instance, in the case of spatially embedded graphs, the $\Gamma$ matrix can represent the distance between each possible pair of nodes. A matrix $\Gamma$ can also quantify homophily effects between nodal attributes, such as the membership to the same community. Following the same steps as above, we get an equivalent expression to eq.(\ref{e-fep-d-approx}):
\begin{equation}\label{e-av_spin_den_ed_cov}
    \frac{d}{dt}\av{\sigma_{ij}}_t \sim -2\mu\av{\sigma_{ij}}_t + \frac{\varphi}{2L}\gamma_{ij}\big[1-\av{\sigma_{ij}}^2_t\big]\ ,
\end{equation}
A solution of the same form as eq.(\ref{e-fep-sol}) can be written for each spin average $\av{\sigma_{ij}}$. However, each eq.(\ref{e-av_spin_den_ed_cov}) depends now on the dyad-specific value $\gamma_{ij}$, by consequence the same is true for the fixed points of the dyad dynamics.

\subsection{Distance-like $F$}\label{ss-dist-F}
We finally discuss an exploitation scheme driven by a distance-like $F$ metric. In this case, an optimal state is explicitly encoded as the global maximum of $F$, which corresponds to a zero distance between a set of graph statistics $\bm{x}(G)$ and the target statistics $\bm{x}^*$, i.e.,
\begin{equation}
    F(G)\propto -\lVert \bm{x}(G)-\bm{x}^*\rVert_2^2\ ,
\end{equation}
where $\lVert\cdot\rVert_2$ indicates an $l^2$-norm. In particular, we consider the simple such case in which the $x(G)=\sum_{ij}a_{ij}$ is one-dimensional and corresponds to the edge count of a graph, the optimal (target) number being $E^*$:
\begin{equation}\label{e-mrg}
    F(G) = -\frac{1}{L^2}\Big( \sum_{i<j}a_{ij}-E^*\Big)^2,
\end{equation}
where $0<E^*<L$. We can follow the same steps as in sec. \ref{s-epen}. The expression analogous to eq.(\ref{e-fep-d}) is
\begin{equation}\label{e-mrg-1}
\begin{split}
    \frac{d}{dt}\av{\sigma_{ij}}_t = -2\mu\av{\sigma_{ij}}_t - \frac{\varphi}{L^2}&\Bigg[\Bigg(\frac{L}{2}-E^*\Bigg)\sum_{k<l}\big(\av{\sigma_{ij}\sigma_{kl}}_t-\av{\sigma_{ij}}_t\av{\sigma_{kl}}_t\big)\ +\\
    &+ \frac{1}{4}\sum_{\substack{k<l,m<n\\(kl)\ne(mn)}} (\av{\sigma_{ij}\sigma_{kl}\sigma_{mn}}_t-\av{\sigma_{ij}}_t\av{\sigma_{kl}\sigma_{mn}}_t)\Bigg]\ .
\end{split}
\end{equation}
We generalize the decoupling approximation in sec. \ref{s-epen} and define it as the regime in which all moments factorize: 
\begin{equation}
    \av{\sigma^{(1)}\sigma^{(2)}\dots\sigma^{(k)}}\sim\av{\sigma^{(1)}}\av{\sigma^{(2)}}\dots\av{\sigma^{(k)}} \ ,
\end{equation}
where the left-hand side of the equation contains no repeated spin variables. Eq.(\ref{e-mrg-1}) becomes:
\begin{equation}\label{e-mrg-approx}
    \frac{d}{dt}\av{\sigma_{ij}}_t = -2\mu\av{\sigma_{ij}}_t - \frac{\varphi}{L^2}\bigg[1-\av{\sigma_{ij}}_t^2\bigg]\Bigg(\frac{L}{2}-E^* + \frac{1}{2}\sum_{\substack{k<l\\(kl)\ne(ij)}} \av{\sigma_{kl}}_t  \Bigg)\ ,
\end{equation}
valid again for $L\epsilon\ll1$, where $\epsilon$ is the order of magnitude of the spin-covariances.
Finally, under the hypothesis of using the same initial condition for the dynamics of all dyads, then we can approximate the last sum $\sim (L-1)\av{\sigma_{ij}}_t$. All dynamical equations are then completely decoupled and an equation of the same form as eq.(\ref{e-mrg-1}) can be written for the magnetization $m_t$
\begin{equation}\label{e-ansol}
\dot{m}_t = -2\mu m_t-\frac{\varphi}{L^2}(1-m_t^2)\Bigg[\frac{L-1}{2}m_t+\frac{L}{2}-E^*\Bigg] \ .
\end{equation}

\begin{figure}[htbp]
  \begin{minipage}{0.5\textwidth}
    \centering
    \includegraphics[width=\textwidth]{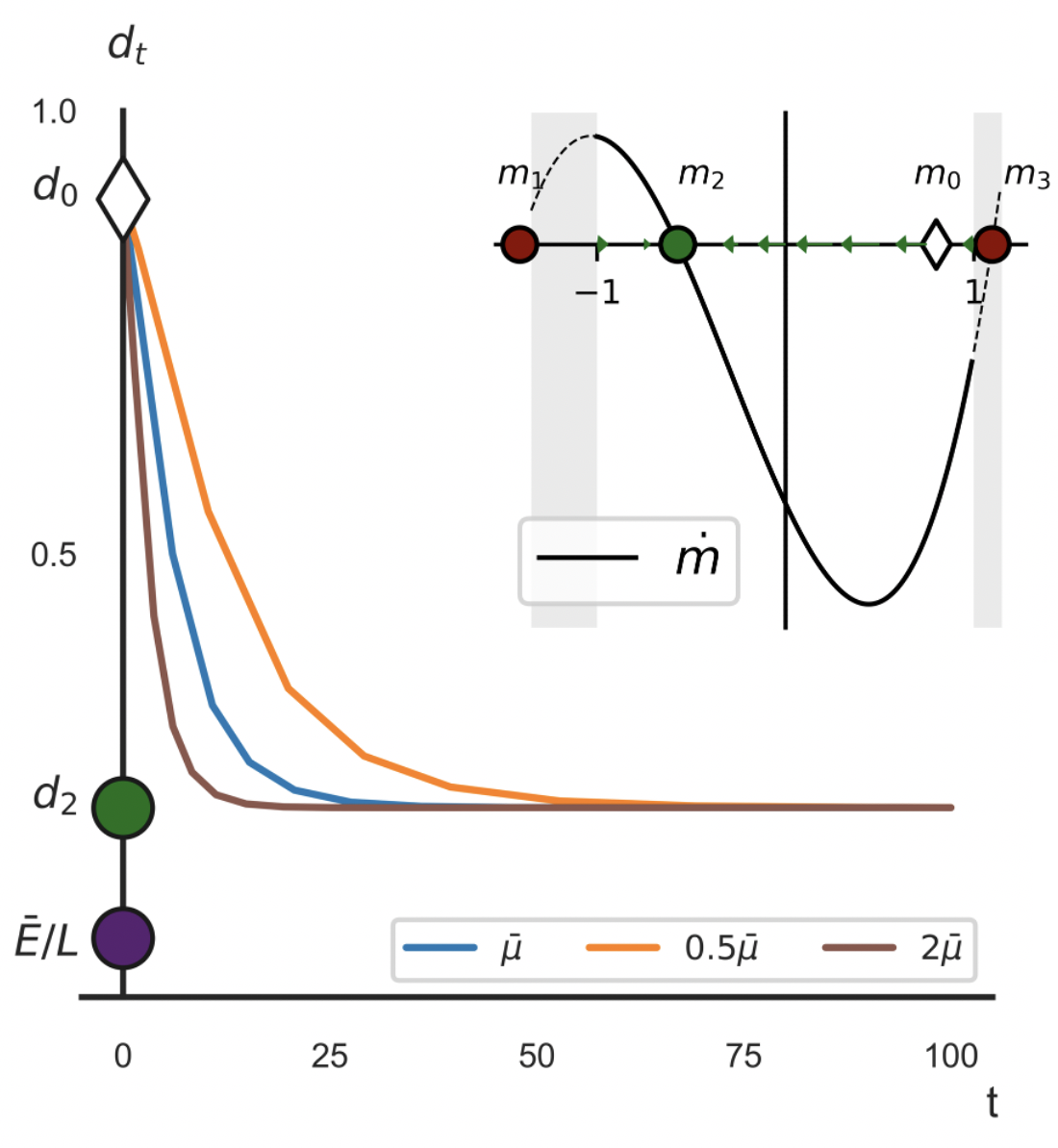}
  \end{minipage}%
  \hfill
  \begin{minipage}{0.45\textwidth}
    \caption{Analytical solution of the exploration-exploitation dynamic based on eq.(\ref{e-ansol}), decoupling approximation. Here $N=10$ ($L=45$), $\bar{\mu}=1/L, \bar{\rho}=5\times10^2$, $d_0=0.9$, $E^*=3$. Inset: plot of $\dot{m}=f(m)$, where $f$ is specified by eq.(\ref{e-ansol}). The dynamic has in this case $3$ fixed points, labelled as $m_1,m_3$ (red, unstable) and $m_2$ (green, stable), found by solving numerically the equation $f(m)=0$. Main: dynamic of the average graph density $d_t$ (numerical solution). Similarly to the case discussed in sec. \ref{s-epen}, the value of the asymptotic state $d_2$ solely depends on $L, \bar{\rho}$ (held fixed) while the rapidity of the approach to it depends on the exploration rate $\mu$. }
    \label{f-mrg}
    \vspace{1em}
  \end{minipage}
\end{figure}

The integration of the above expression with boundary condition $m_{t_0}=m_0$ completely solves the one-dimensional system dynamic and can be done numerically, fig.(\ref{f-mrg}). We find the same qualitative behaviour as in sec. \ref{s-epen}, with the solution approaching an asymptotic value that strikes a balance between the influx of mutations - driving the system towards $d=1/2$ - and the strength of the exploitation - steering the system towards $E^*/L$. The rapidity of the approach to the asymptotic density value $d_{\infty}$ depends on $\mu$. 

\section{Population-based simulations}\label{s-sim}
We further exploit the analogy with evolutionary models mentioned in sec. \ref{s-darwin} to design population-based simulations for eq.(\ref{e-me}) \cite{zanini2012,mauri2021}. The current implementation has been coded in \texttt{Python 3.9.7} and is available on Github \cite{dichio2023git}.

In this study, we consider graphs that are unweighted ($a_{ij}=0,1$), undirected ($a_{ij}=a_{ji}$), with no self-loops ($a_{ii}=0$). Therefore, if $N$ is the (fixed) number of nodes, each graph can be represented by a string of $L=N(N-1)/2$ binary values $0100\dots01$ and the configuration space is the $L$-dimensional hypercube $\{0,1\}^L$. Instead of tracking the dynamic of a single instance of the system (individual) in the configuration space, we design a population-based framework where we track simultaneously the evolution of many samples of the probability distribution. More specifically, if $G\in\mathcal{G}$ is a graph configuration and $n\in\mathbb{N}$ is the number of times $G$ is observed in the population, we call the couple $(G,n)$ a clone. At any time $t$, a population $\mathcal{P}(t)=(\bm{G}(t),\bm{n}(t))$ is defined as the set of existing clones. The total number of individuals $\sum_{\alpha}n_{\alpha}(t)=M$ is held fixed while the total number of clones $M_c(t)\le M$ fluctuates. At each simulation step $\Delta t$ the population undergoes the processes of mutations (exploration) and selection (exploitation) as follows:

\begin{figure}
    \centering
    \includegraphics[width=.8\columnwidth]{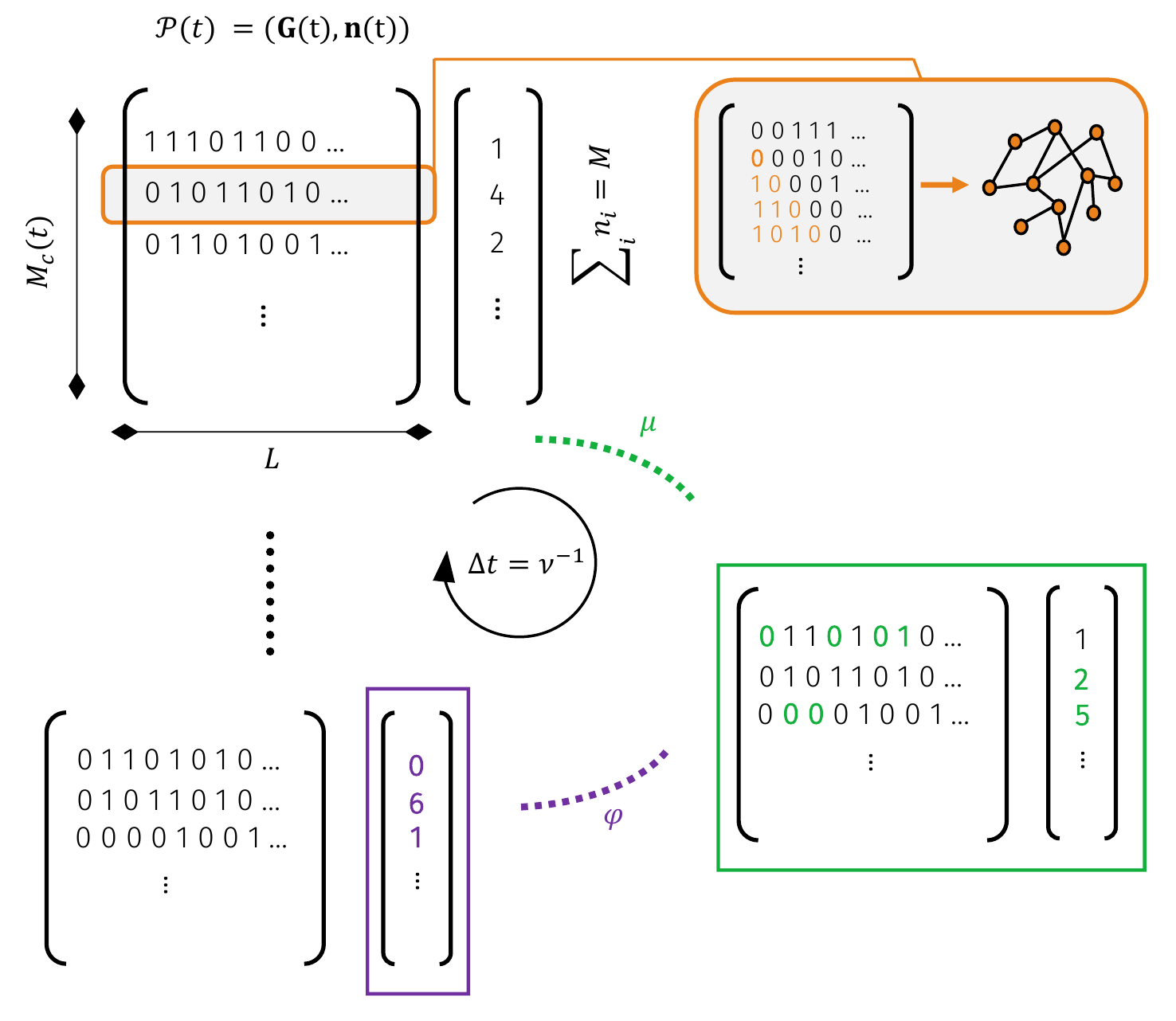}
    \caption{A simulation step of population-based simulations for eq.(\ref{e-me}). A population $\mathcal{P}(t)$ at time t consists of matrix $\bm{G}(t)$ with dimensions $M_c(t)\times L$ and a vector $\bm{n}(t)$ of length $M_{c}(t)$ (top left). Each row of $\bm{G}(t)$ corresponds to a unique graph configuration, and in particular to the vectorized lower triangular matrix of the corresponding adjacency matrix (top right). First, the population undergoes mutations, with rate $\mu$ per each dyad (green, bottom right), here bit-flips. Consequently, both the matrix $\bm{G}$ and the counts $\bm{n}$ change. As new configurations are found, the number of clones $M_c$ increases. Second, selection is implemented, tuned by the parameter $\varphi$ (violet, bottom left). Selection only affects the vector of counts $\bm{n}$, those graphs that end up with $n=0$ are removed from $\bm{G}$ before the next time-step. The total number of individuals $M$ and time interval $\Delta t$ are free internal parameters of the simulations.
    }
    \label{f-sim}
\end{figure}

\begin{itemize}
    \item \emph{Exploration}. Each possible dyad of each existing graph in the population mutates with probability $1-e^{-\Delta t\mu}\sim\Delta t\mu$ for $\Delta t\mu\ll1$. The mutation (exploration) rate is uniform across dyads. Using spin-variables eq.(\ref{e-spdy}), in the case of bit-flip mutations we have $\sigma_{ij}\rightarrow-\sigma_{ij}$. If edges are only allowed to appear, $\sigma_{ij}\rightarrow|\sigma_{ij}|=1$.
    \item \emph{Exploitation}. Prior to selection, the $F$ metric of all existing graph configurations $G_{\alpha}$ is computed. The counts $\bm{n}$ are then updated by performing M independent draws from a multinomial distribution where each $G_{\alpha}$ is chosen with probability
    \begin{equation} \label{e-pas}
        p_{\alpha} =  n_{\alpha}e^{\Delta t \varphi F(G_{\alpha})}/\sum_{\beta}n_{\beta}e^{\Delta t \varphi F(G_{\beta})}\ ,\qquad \alpha\in 1,\dots,M_c(t) \ .
    \end{equation}
\end{itemize}
    
An illustration of the workflow of a simulation step can be found in fig.(\ref{f-sim}). A pseudo-code for the simulations is the following:

\smallskip

\begin{algorithmic}
\State $\mathcal{P}(0) = (\bm{G}(0),\bm{n}(0))$
\State $t = 0$
\While{$t< T$} 
\State \emph{Exploration:} $\sigma_{ij}\rightarrow-\sigma_{ij}$ with probability $\Delta t\mu\ $ $\forall (i,j)$, $\forall G_{\alpha}$\;
\State update $\mathcal{P}^* = (\bm{G}^*,\bm{n}^*)$
\State compute $F_{\alpha}= F(G_{\alpha}^*) \ \forall \ G_{\alpha}^*\in \bm{G}^*$
\State \emph{Exploitation:} $M$ draws $\sim$ multinom. distrib. with $p_{\alpha} = n_{\alpha}^*e^{\Delta t \varphi F(G_{\alpha}^*)}/\sum_{\beta} n_{\beta}^*e^{\Delta t \varphi F(G_{\beta}^*)}$ $\Rightarrow$ new counts $\bm{n}^{**}$\;
\State set $\mathcal{P}(t)= (\bm{G}^*,\bm{n}^{**})$
\State $t=t+\Delta t$
\EndWhile
\end{algorithmic}
\smallskip

Such a simulation framework has hence two internal free parameters: the total number of samples $M$ and the simulation step $\Delta t$ (or, equivalently, its inverse $\nu=\Delta t^{-1}$). Tab.(\ref{t-params}) summarizes all the parameters needed for a single run. A distinction is made between (i) the structural (struct) parameters $N,T$, which set the dimension of the configuration space and the length of the time window (ii) internal simulation (sim) parameters $M, \nu$, mentioned above and (iii) the dynamical (dyn) parameters $\mu, \rho$, which tune the system dynamic.

\begin{table}[htbp]
  \centering
  \caption{Parameters of simulations for EE graph dynamics. Our computational framework has six degrees pf freedom, which we group by color: structural parameters (top), internal degrees of freedom (middle) and parameters of the dynamics (bottom).}
  \begin{tabularx}{.4\textwidth}{>{\centering\arraybackslash\hsize=.5\hsize}X>{\raggedright\arraybackslash\hsize=1.5\hsize}X}
    \textbf{Parameter} & \textbf{Description} \\
    \toprule
    \rowcolor[HTML]{DFD7BF}
    $N$  &  number of nodes  \\
    \rowcolor[HTML]{DFD7BF}
    $T$  &  time window span  \\
    \addlinespace[3pt]
    \rowcolor[HTML]{F2EAD3}
    $M$  &  population size \\
    \rowcolor[HTML]{F2EAD3}
    $\nu$  &  inverse time step $\Delta t^{-1}$ \\
    \addlinespace[3pt]
    \rowcolor[HTML]{F5F5F5}
    $\mu$  & exploration rate \\
    \rowcolor[HTML]{F5F5F5}
    $\rho$  & functional pressure $\varphi/\mu$ \\
    \bottomrule
  \end{tabularx}
  \label{t-EE-params}
\end{table}

\begin{figure}
  \centering
    \subfloat[\label{f-comptime-M}]{\includegraphics[width=0.49\linewidth]{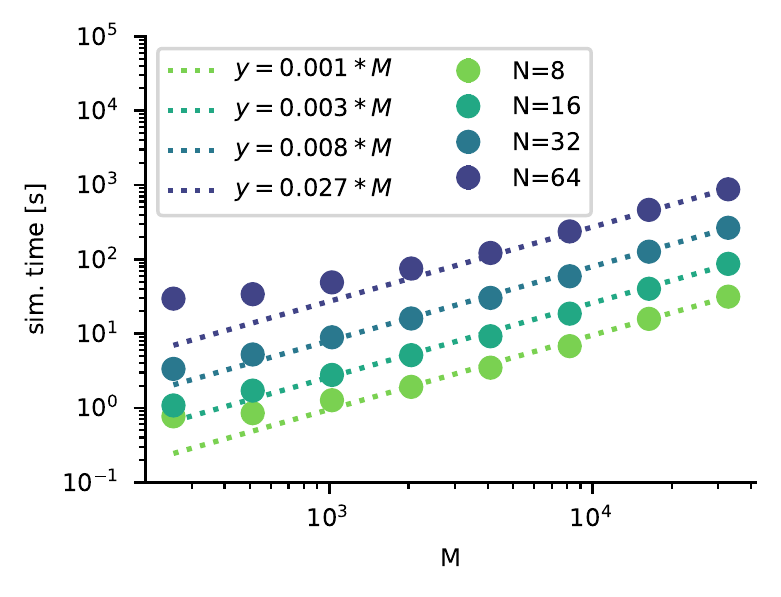}
    }\hfill
    \subfloat[\label{f-comptime-N}]{\includegraphics[width=0.49\linewidth]{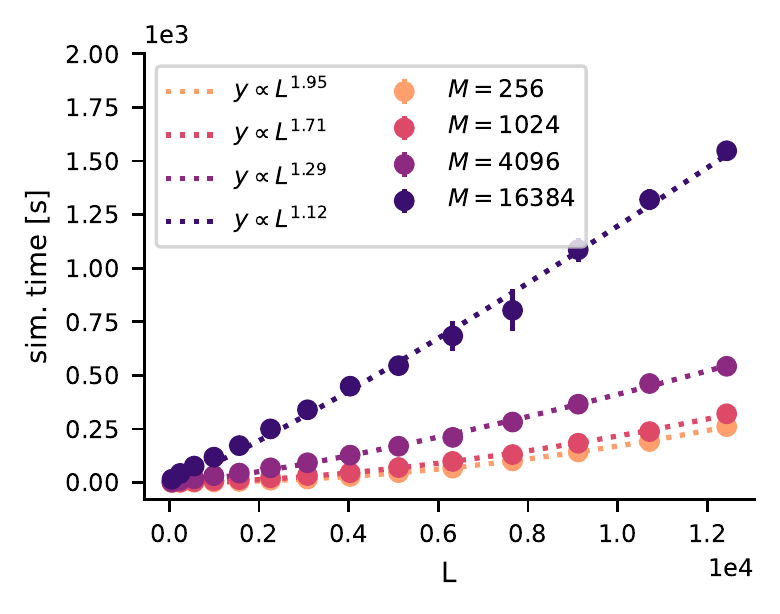}
    }
  \caption{Simulation time as a function of the number of individuals $M$ and $L=N(N-1)/2$, where $N$ is the number of nodes. Each dot is the average value obtained from $10$ simulation runs, error bar are indicated (when visible). Here $T=100, \nu=1, \mu=1.0\times10^{-4},\rho=5$, an $F$ metric as in sec. \ref{s-epen} is used. (a) The simulation time is linearly dependent on the number of individuals $M$. Dotted lines are the result of a linear fit with the curve $y=a*M$, where $a$ is a parameter. A log-log scale has been used. (b) The simulation time for this problem has a mild exponential dependence on $L$. Dotted lines are the result of a fit with the curve $y=a*L^{b}$, where $a,b$ are parameters. In the limit of large $M$, we find $\beta\sim1$, indicating an approximately linear dependence on $L$. The dependence on $L$ ($N$) however, strongly depends on the complexity of the $F$ metric.}
  \label{f-comptime}
\end{figure}

The simulation time has an obvious linear scaling with the inverse time interval $\nu$, since one run of simulation involves repeating the above steps a number $\nu T$ of times, where $T$ is the final time. In fig.(\ref{f-comptime-M}) we show that a there is a liner dependence from the number of individuals in a population $M$, too. Each simulation steps in fact involves mutating $M$ individuals independently and (in the worst case) evaluating the $F$ metric of $M$ different individuals. In fig.(\ref{f-comptime-N}) we show that, in the limit of large $M$, a linear dependence is also found on $L$. This latter result however is strongly dependent on the choice of the $F$ metric, which represents the potential bottleneck of the simulation framework here presented.
Additional details can be found in the documentation available at \cite{dichio2023git}.

\subsection{Distance-like $F$, simulated}
Let us briefly re-consider the distance-like $F$ metric eq.(\ref{e-mrg}), discussed in sec. \ref{ss-dist-F}. We can set-up the corresponding simulations, use them to showcase the essential features of a single simulation run and compare it to the analytical approximation eq.(\ref{e-mrg-approx}). 

At each given point in time $t$, we can use the graph population $\mathcal{P}(t)=(\bm{G}(t),\bm{n}(t))$ to approximate the distribution of any any observable $O:\mathcal{G}\rightarrow\mathbb{R}$ by 
\begin{equation}\label{e-emp-dis}
    P(O,t) = \frac{1}{M}\sum_{\alpha=1}^{M_c(t)} n_{\alpha}(t)\ \delta[O-O(G_{\alpha}(t)]\ ,
\end{equation}
where $\delta$ is the Dirac-delta function ($\int dO \ \delta(O) = 1$). By consequence, the expected value of $O$ at time $t$ is
\begin{equation}\label{e-emp-av}
    \av{O}_t\sim\frac{1}{M}\sum_{\alpha=1}^{M_c(t)} n_{\alpha}(t) \ O(G_{\alpha}(t))\ .
\end{equation}

\begin{figure}[ht!]
\includegraphics[width=.8\columnwidth]{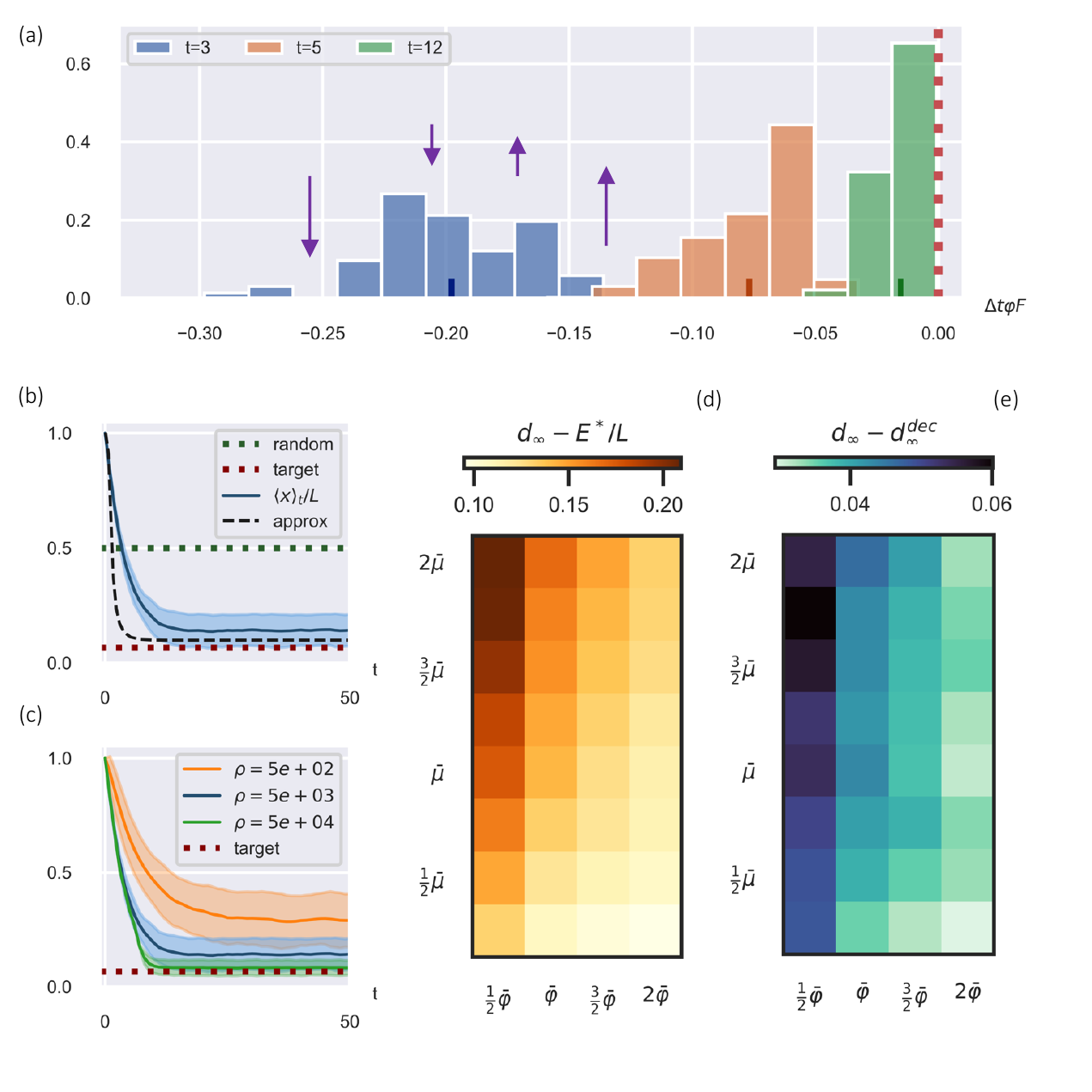}
\caption{\label{f-mrg-sim}
Exploration-exploitation (EE) dynamic as defined by eq.(\ref{e-me}, \ref{e-mrg}), $N=10$, $\bar{\mu}=1/L$, $\bar{\rho}=5\times10^{3}$ ($\bar{\varphi}=\bar{\rho}\bar{\mu}$), $T=50$, $\Delta t = 1/\nu = 10^{-2}$, $E^*=3$, $\Delta t =1$, $M=4096$.
(a) Distribution of $F$ values at different $t$, the corresponding $\av{F}_t$ are indicated. At each $t$, exploitation increases the probability of those graphs with higher $F$ than the ensemble average (violet arrows). As a result, the distribution moves towards higher $F$ values, until the max is reached (red dotted line). (b) The average graph density $\av{x}/L$ evolves towards the target density $E^*/L$, until a stationary value $\av{x}_{\infty}/L$ is reached - shaded area indicates the $95\%$ confidence interval, black dashed line indicates the analytical solution from eq.(\ref{e-ansol}). (c) The value $\av{x}_{\infty}/L$ and the rapidity of the approach to it depends on the functional pressure $\rho$, it approaches $E^*/L$ for increasing exploitation rate $\varphi$ and decreasing exploration rate $\mu$. (d) The distance from the target increases for increasing exploration rate $\mu$ and decreasing selection rate $\varphi$. (e) Difference between $d_{\infty}$ as in (d) and the corresponding value $d_{\infty}^{dec}$ decoupling approximation eq.(\ref{e-mrg-approx}). The $d_{\infty}^{dec}$ is always closer to the target than $d_{\infty}$, the difference between the two vanishing for increasing $\varphi$ (where both approach $E^*/L$), as expected.
}
\end{figure}

Fig.(\ref{f-mrg-sim}a) illustrates the EE dynamic in the space of the $F$ values. At any given time $t$ the $F$-distribution eq.(\ref{e-emp-dis}) is subject to two opposing forces. On the one hand, the influx of random dyadic mutations drives the distribution towards the $F$ of the maximally random state, an Erd\H{o}s-Rényi random graph with $p=0.5$ \cite{albert2002}. On the other hand, the exploitation term increases(decreases) the probability of those $F$ which are above(below) $\av{F}_t$, resulting in a net movement of the distribution towards higher $F$ values. This is analogous to a mechanism of \emph{adaptation} in evolutionary dynamics.

In fig.(\ref{f-mrg-sim}b) we show the resulting simulated dynamic of $d_t=\av{x}_t/L$ and the corresponding confidence interval -- whose calculations are based on eq.(\ref{e-emp-av}). A stationary value is reached where the strengths of the exploration-exploitation drivers are balanced. The value of this stationary point and the speed at which it is approached depend on the functional pressure $\rho$, fig.(\ref{f-mrg-sim}c). In particular, the higher $\rho$ the closer will be the asymptotic solution to the optimal value, fig.(\ref{f-mrg-sim}d). Finally, in fig.(\ref{f-mrg}e) we compare the asymptotic values $d_{\infty}$ resulting from simulations with the corresponding values $d_{\infty}^{dec}$ obtained numerically under decoupling approximation, eq.(\ref{e-mrg-approx}). We show evidence that $d_{\infty}^{dec}$ (decoupling approximation) always lies closer to the target value $E^*/L$ than $d_{\infty}$, the difference between the two vanishing for increasing exploitation strength $\varphi$. We conclude that, for fixed $N$, the decoupling approximation agrees qualitatively everywhere in the parameter space with simulations, see fig.(\ref{f-mrg-sim}b), quantitatively for large $\varphi$. 

\section{Exponential random graph (ERG) models in a nutshell} \label{s-ERGM}
Exponential random graph (ERG) models in this work serve the purpose of providing a parsimonious characterization of a given graph $G^*$. They belong to broad class of maximum entropy (maxent) inference methods \cite{jaynes1957}, and are designed for graph data. We here briefly illustrate the essential features of the method, further details can be found in \cite{cimini2019,dichio2022} and references therein.

Let $G^*\in\mathcal{G}$ be a single realization of the system under investigation. Suppose the existence of $r$ \emph{sufficient statistics} $\bm{x}:\mathcal{G}\rightarrow\mathbb{R}^r$ for $G^*$, i.e., that characterize it. According to the maximum entropy principle (MEP), the most unbiased probability density function $P(G)$ consistent with the available information $\bm{x}(G^*)$ is obtained by maximizing the Shannon information entropy $\mathcal{S}$
\begin{equation}\label{eSE}
    \mathcal{S}[P]=-\sum_{G\in\mathcal{G}}P(G)\log P(G)
\end{equation}
while at the same time imposing the normalization $\sum_{G\in\mathcal{G}} P(G) = 1$ and the soft constrains:
\begin{equation}\label{e-constraints}
    \sum_{G\in\mathcal{G}}\bm{x}(G)P(G) = \bm{x}(G^*) \ . 
\end{equation}
This maximization problem is easily solved by introducing a set of Lagrange multipliers $\bm{\theta}\in\mathbb{R}^r$, and the result is the ERG probability distribution:
\begin{equation}\label{eERGM}
    P(G|\bm{\theta})= \frac{e^{\bm{\theta}\cdot\bm{x}(G)}}{\sum_{\Tilde{G}\in\mathcal{G}}e^{\bm{\theta}\cdot\bm{x}(\Tilde{G})}}\ ,
\end{equation}
where the $\bm{\theta}$ are set so to satisfy eq.(\ref{e-constraints}). The parameters $\bm{\theta}$ can be interpreted by quantifying their effect on the likelihood that an edge exists between any pair of nodes $i,j$ in the graph. Let $P(a_{ij}=1|G_{\backslash ij},\bm{\theta}^*)$ be the probability of an edge within the dyad $(i,j)$, given the rest of the graph $G_{\backslash ij}$. We have
\begin{equation}\label{e-log-odds}
    \log\ \frac{P(a_{ij}=1|G_{\backslash ij},\bm{\theta}^*)}{P(a_{ij}=0|G_{\backslash ij},\bm{\theta}^*)} = \bm{\theta}^*\cdot \bm{\Delta x}(G_{ij})\ ,
\end{equation}
where $\bm{\Delta x}(G_{ij})$ is the vector of \emph{change statistics}. Defining $G_{+ij} = \{a_{ij}=1,G_{\backslash ij}\}$ and $G_{-ij} = \{a_{ij}=0,G_{\backslash ij}\}$, the $\alpha$-th element of the change statistics is $\Delta x_{\alpha}(G_{ij}) = x_{\alpha}(G_{+ij}) - x_{\alpha}(G_{-ij})$.  The dyadic interpretation \eqref{e-log-odds} of the parameter $\theta_{\alpha}$ is the change in the log probability of a graph, resulting from switching from $G_{-ij}$ to $G_{+ij}$ (i) per unit increase of the corresponding statistic $\Delta x_{\alpha}(G_{ij})=1$, and (ii) holding fixed the cumulative effect of the other statistics $\sum_{\beta\ne\alpha} \theta_{\beta}\Delta x_{\beta}(G_{ij})$. The interpretation of the ERG parameters is therefore based on a characterisation of the ensamble distribution $P(G|\bm{\theta}^*)$. This because, by construction eq.\eqref{e-constraints}, the properties of the latter reflect those of the original graph. Large positive (negative) values of the parameter $\theta_{\alpha}$ indicate an over- (under-) representation in the original graph of the corresponding $x_{\alpha}$, with respect to the null expectation -- i.e., $\av{x_{\alpha}}$ of an ERG model with $\theta_{\alpha}=0$ and unaltered $\bm{\theta}_{\backslash \alpha}$ \cite{dichio2022,dichio2023phd}. % C.3 

Given a graph $G^*$, the parameters $\bm{\theta}$ of an ERG model can be inferred from data. In theory, this is achieved by maximizing the log-likelihood 
\begin{equation}\label{eMAXLIK-EST}
    \bm{\theta}^* = \argmax_{\bm{\theta}} \ \log P(G^*|\bm{\theta})\ .
\end{equation}
In practice, the latter computation is hampered by the evaluation of the partition function, i.e., the denominator of eq.(\ref{eERGM}). Nevertheless, the inference problem can be solved numerically by resorting to standard approximation schemes for exponential probability distributions. In particular, Markov chain Monte Carlo maximimum likelihood estimation (MCMC-MLE) iteratively explores the parameter space $\mathbb{R}^r$ looking for the set of parameters $\bm{\theta}^*$ that solves the maximization problem eq.(\ref{eMAXLIK-EST}). This is done by exploiting eq.(\ref{e-log-odds}) to sample graphs from eq.(\ref{eERGM}) with MCMC routines and bypassing the computation of the partition function, an idea dating back to the $90$s \cite{geyer1992}. A number of packages have been developed in recent years to perform such inference task, the implementation used in this work is the \texttt{ergm} package of the \texttt{statnet} suite \cite{krivitsky2022}, written in \texttt{R} language. The inferred parameters $\bm{\theta}$ serve the purpose of assigning weights to each statistic and quantifying their contribution towards the generation of the observed graph.

\subsection{ERG statistics}\label{ss-ERGM-stats}
\begin{figure}
    \centering
    \subfloat[\label{f-gwd-gwesp}]{\includegraphics[width=0.45\linewidth]{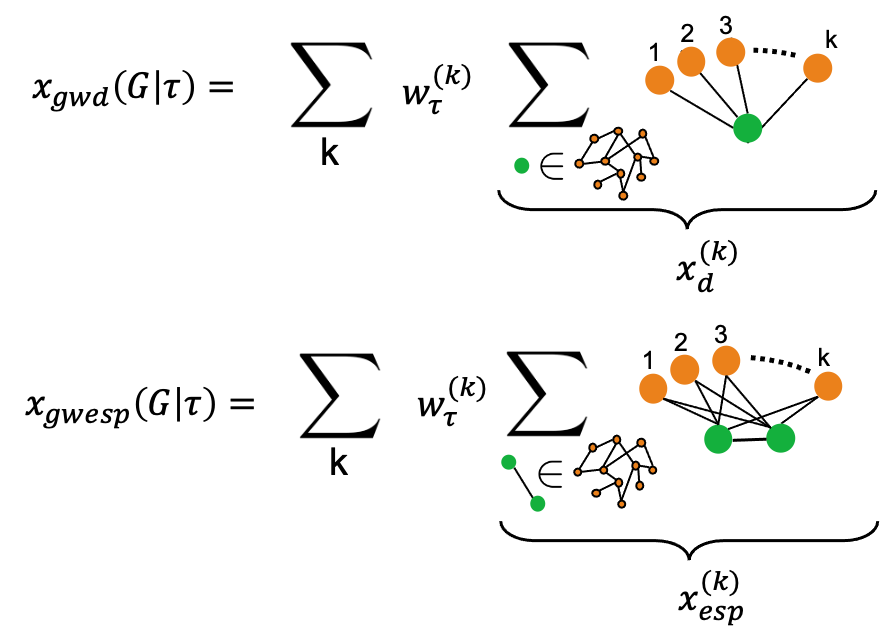}
    }\hfill
    \subfloat[\label{f-gwdecay}]{\includegraphics[width=0.5\linewidth]{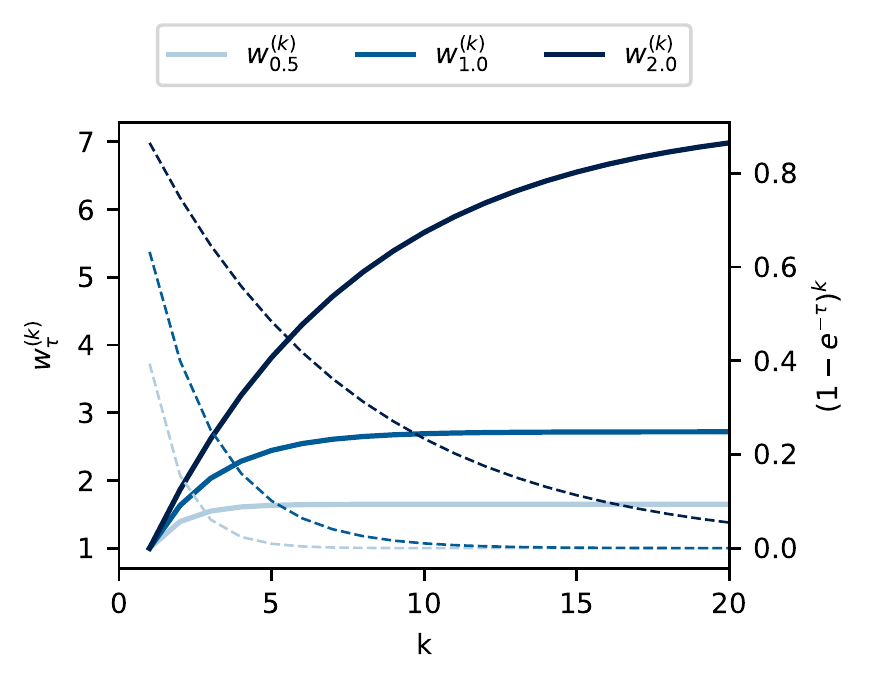}
    }
  \caption{Geometrically weighted statistics. (a) An illustration of the \texttt{gwd} and \texttt{gwesp} statistics, as defined in eq.(\ref{eTERM-GWD}-\ref{eTERM-GWESP}). Both have the structure of a linear combination of a distribution (degrees, edgewise shared partners), the weights $w_{\tau}^{(k)}$ are defined in eq.(\ref{e-gweigths}). (b) Weights $w_{\tau}^{(k)}$ (solid lines) and geometric decaying factors (dashed lines), as they appear in eq.(\ref{e-logodds-c}). Consider as an example the case of $x_{gwesp}$, in the case $\theta_{gwesp}>0$. Adding a new partner to a pair of connected nodes $k\rightarrow k+1$ always leads to an increase in the probability associated to the graph, eq.(\ref{e-logodds-c}), since eq.(\ref{e-gweigths}) is a monotonically increasing function of $k$. However, this increase diminishes and eventually vanishes for increasing $k$.  }\label{f-curvedstats} % C.3
\end{figure}

The starting point of an ERG model -- as well as of all maxent inferential methods -- is the choice of statistics, which is up to the modeller. Choosing a set of statistics entails making an hypothesis about the effects that are \emph{relevant} for the system. For a given choice of the statistics, the ensemble eq.(\ref{eERGM}) - whose parameters are inferred from data - defines a minimal model in which only the information represented by $\bm{x}$ is accounted for, and is otherwise maximally unbiased.
A huge variety of statistics $x:\mathcal{G}\rightarrow\mathbb{R}$ has been proposed and might be included \cite{krivitsky2022}. In this work, we have used in particular two kind of statistics. The first, edge covariates, have already been introduced in eq.(\ref{e-edgecov}). We here briefly describe the second, the so-called \emph{curved statistics}, fig.(\ref{f-gwd-gwesp}). 

Let us consider for instance the geometrically weighted degree \texttt{gwd}. It is based on the graph degree distribution and is able to account, e.g., for an over/under representation of hubs in a graph, while at the same time avoiding the so-called \emph{degeneracy} problem of simplistic ERG models \cite{schweinberger2020}. Formally, 
\begin{equation}\label{eTERM-GWD}
    x_{gwd}(G|\tau) = e^{\tau}\sum_{k=1}^{N-1} \Big\{1-\big(1-e^{-\tau}\big)^{k}\Big\}x_{d}^{(k)}(G)\ ,
\end{equation}
where $\tau>0$ is a decay parameter and $x_d^{(k)}(G)$ is the number of nodes in the graph $G$ with degree $k$. It is a linear combination of the degree distribution, where the linear coefficients 
\begin{equation}\label{e-gweigths}
    w^{(k)}_{\tau} = e^{\tau}\Big\{1-\big(1-e^{-\tau}\big)^{k}\Big\}
\end{equation}
are based on the geometric series $(1-e^{-\tau})^{k}$. In order to interpret the effect accounted by eq.(\ref{eTERM-GWD}), we can reason as follows \cite{hunter2007,dichio2022}. Let us consider a single edge between the nodes $i,j$. As a consequence of this addition, the degrees of both the incident nodes increase, let us focus on one such increase $k\rightarrow k+1$, 
using eq.(\ref{eERGM}) one finds:
\begin{equation}\label{e-logodds-c}
    \log \frac{P(a_{ij}=1|G_{\backslash ij},\bm{\theta}^*)}{P(a_{ij}=0|G_{\backslash ij},\bm{\theta}^*)} \propto \theta_{gwd} (1-e^{-\tau})^{k}  \ .
\end{equation}
The role of the ERG parameter $\theta_{gwd}$ is the same as described in eq.(\ref{e-log-odds}). Let us consider the case $\theta_{gwd}>0$, which corresponds to the scenario we found in \ms. Eq.(\ref{e-logodds-c}) implies that the relative advantage in adding an edge to a node with degree $k$ decreases geometrically with $k$, fig.(\ref{f-gwdecay}). The rapidity of the decay is controlled by $\tau$, a large $\tau$ resulting in a slow decay. It should be noted that, without the geometric decay, an MCMC dynamic based on eq.(\ref{e-logodds-c}) would end up in a fully connected graph (\emph{degeneracy}), making the estimation impossible. Therefore, for $\theta_{gwd}>0$, a bias towards high-degree nodes is present, but thanks to the geometric weights it attenuates and eventually disappears for increasing $k$ - where $w_{\tau}^{(k)}$ reaches a plateau and $(1-e^{-\tau})^{k}\sim0$. Thus, the system is not steered into an unrealistic fully-connected graph state.

In addition to the \texttt{gwd}, we also introduce the geometrically weighted shared partner distribution:
\begin{equation}\label{eTERM-GWESP}
    x_{gwesp}(G|\tau) = e^{\tau}\sum_{k=1}^{N-2} \Big\{1-\big(1-e^{-\tau}\big)^{k}\Big\}x_{esp}^{(k)}(G)\ ,
\end{equation}
where $x^{(k)}_{esp}(G)$ is the number of pairs of connected nodes that share exactly $k$ neighbors. The latter statistic is employed to capture an over/under representation of triads in the network, for $\theta_{gwesp}\gtrless0$. As in the previous case the use of geometric weights helps to prevent this tendency from implying an unrealistic fully connected graph.

\section{\emph{C.elegans} connectome(s)}
\subsection{Overview}
The \ce\ connectome - i.e., the wiring diagram of neural connections within the worm nervous system -, has garnered significant attention within the field of neuroscience, due to its complete mapping and simplicity \cite{bentley2016,cook2019}. It provides valuable insights into the relationship between neural networks and behavior, making it an ideal model system for understanding fundamental principles of brain organization and function. 

%% Qualitative : adult
The nervous system of an adult, hermaphrodite \ce\  was first described in the $1986$ by \emph{White et al.} \cite{white1986} and consists of only $302$ cells (neurons), classified into different types based on their morphology, function, and connectivity. These include sensory neurons that detect various environmental cues, interneurons that relay information within the nervous system, motor neurons that control muscle movement and modulatory neurons that release neuromodulators, playing a crucial role in modulating behavior and neural activity. The neurons of the \ce\ are spatially arranged in a stereotypical fashion, the adult body length being $\sim 1\ mm \ $ \cite{skuhersky2022}. A high bilateral symmetry is found, most of the neurons occurring in pairs located along the left and right sides of the body. 
Neurons communicate through different types of cellular junctions. Gap junctions are specialized channels that connect the cytoplasm of adjacent cells, facilitating the rapid exchange of ions, molecules, and electrical signals. Chemical synapses serve as specialised junctions that enable the unidirectional transmission of chemical signals, known as neurotransmitters, from a presynaptic neuron to a postsynaptic neuron or target cell. 

This work focused specifically on the maturation of the \ce\ brain \cite{pathak2020,rapti2020,witvliet2021}. By this, we refer to the process by which the nervous system develops and matures over the worm's life cycle -- the latter includes one embryonic stage (before hatching), four larval stages and one adult stage. At a molecular scale, it involves intricate cellular and molecular events that occur during embryonic and post-embryonic stages. In \ms, we developed a model for synaptogenesis, i.e., the process by which synapses are formed and established in the developing nervous system. It involves the growth and differentiation of neuronal processes, such as axons and dendrites, and the precise alignment and interaction of presynaptic and postsynaptic components. 

A number of online resources are available that provide comprehensive information and tools for studying the \ce\ nervous system, e.g., \href{https://www.wormatlas.org/}{Wormatlas}.

\subsection{Dataset and pre-processing}

\begin{figure}
  \centering
  \includegraphics[width=15cm]{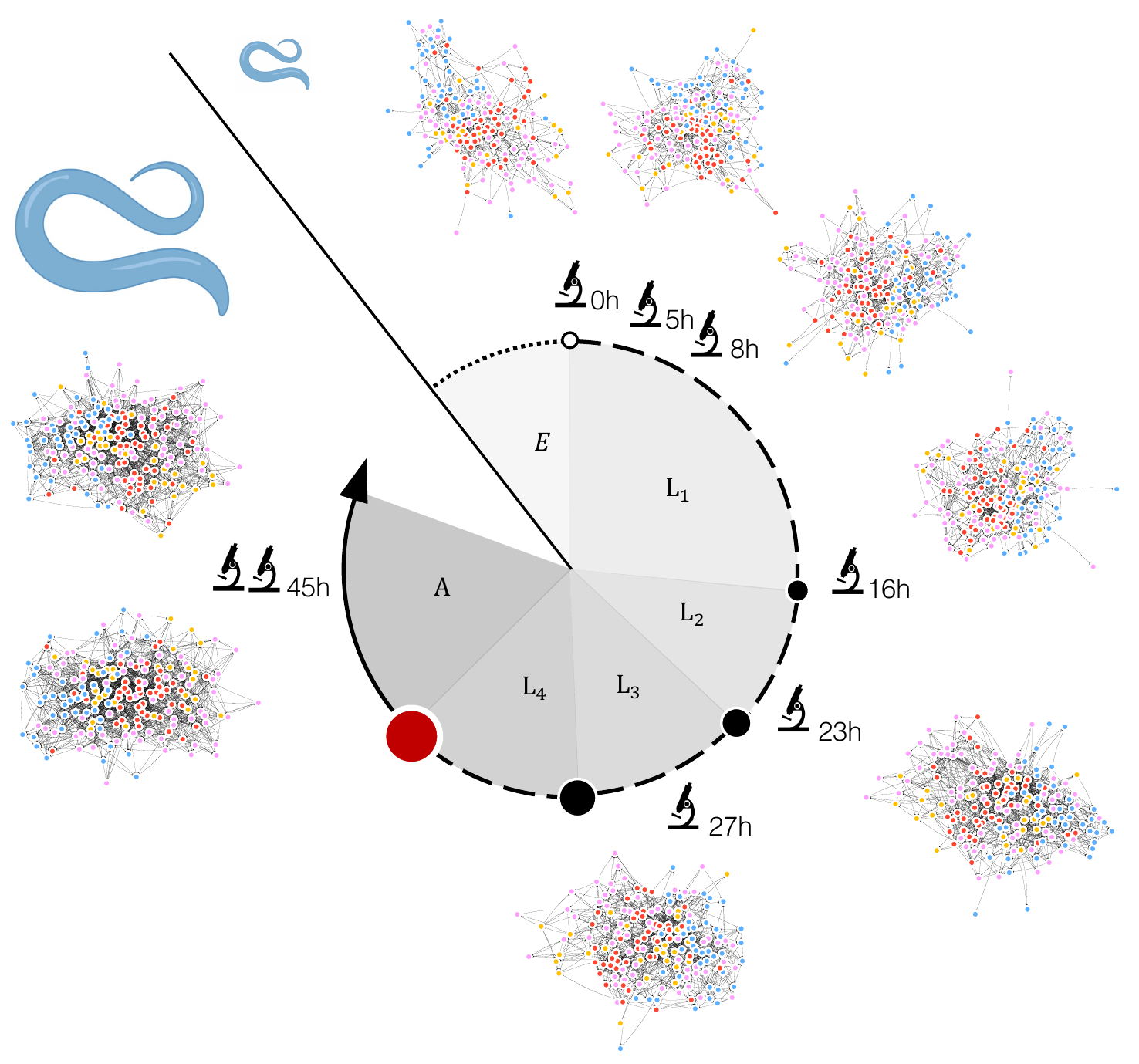}
  \caption{\ce\ brain networks at different developmental ages. The embryonic stage (E, dotted line) terminates with the hatching (birth, $\SI{0}{\hour}$, white circle). The post-embryonic stages include four larval ages ($L_1-L_4$, dashed line) and ends with the onset of adulthood (red circle). The adult stage (A, solid line) lasts about $2-3$ days. The dataset \emph{Witvliet et al.} (2021) \cite{witvliet2021} consists of eight snapshots (microscope icons), including one at birth  $\SI{0}{\hour}$, three $L_1$ $\sim\SIlist{5;8;16}{\hour}$, one $L_2$ $\sim\SI{23}{\hour}$, one $L_3$ $\sim\SI{27}{\hour}$ and two adults, both $\sim\SI{45}{\hour}$. In each network, nodes are individual neuron cells, colored by cell-type: interneurons in red, modulatory in yellow, motor in blue, sensory in pink -- see tab.(\ref{t-neurons}) for a complete list. A directed edge (connection) is placed between two neurons if at least one synapse exists between them.}
  \label{f-snaps}
\end{figure}

In \ms, we modelled the data recently published by \emph{Witvliet et al. (2021)} \cite{witvliet2021}. In the latter, serial-section electron microscopy was used to reconstruct the brains of eight isogenic \ce\ (wild-type N2, hermaphrodites), imaged at different ages and during different post-embryonic stages, see fig.(\ref{f-snaps}). 
The estimation of the age was based on the cell division pattern of the worms at the moment they are selected for imaging. The reconstructed brains, consisting of the nerve ring and ventral ganglion, include $161$ of the total $222$ neurons at birth and $180$ of the total $302$ neurons in adulthood, each cell was unambiguously identified by a code and classified as either sensory, modulatory, or interneuron, tab.(\ref{t-neurons}).
Chemical synapses and gap junctions were manually annotated. The former were fully mapped while the latter were mapped only partially and, for this reason, they were not considered in our analysis.
A directed synaptic connection was defined as a pair of neurons (presynaptic, postsynaptic) connected by at least one chemical synapse. The eight networks of directed synaptic connections were the starting point of our analysis, fig.(\ref{f-snaps}). 

\arrayrulecolor{white} 
\begin{table*}[h!]
  \centering
  \scriptsize
  \setlength{\arrayrulewidth}{2pt}
  \renewcommand{\arraystretch}{1.2}
  \caption{List of the $180$ neurons of the adult \ce\ brain (hermaphrodite, N2), as reported in \cite{witvliet2021}. Interneurons in red, modulatory in yellow, motor in blue, sensory in pink. We have marked with an asterisk$^*$ those neurons that were not present at birth. Each neuron in the worm nervous system is uniquely identified by a code, which consists in two or three letters (or, occasionally, numbers), followed by the position in worm's body D/V (dorsal/ventral), R/L (right/left) \cite{white1986}. The left-right symmetry increases over time and reaches the $\sim90\%$ in the adult brain.}  \label{t-neurons}
  \begin{tabularx}{\columnwidth}{*{14}{|>{\centering\arraybackslash}X}|}
    \hline
    \cellcolor{interneuron} ADAL & \cellcolor{interneuron} ADAR & \cellcolor{interneuron} AIAL & \cellcolor{interneuron} AIAR & \cellcolor{interneuron} AIBL & \cellcolor{interneuron} AIBR & \cellcolor{interneuron} AINL & \cellcolor{interneuron} AINR & \cellcolor{interneuron} AIYL & \cellcolor{interneuron} AIYR & \cellcolor{interneuron} AIZL & \cellcolor{interneuron} AIZR & \cellcolor{interneuron} AVAL & \cellcolor{interneuron} AVAR \\
    \hline
    \cellcolor{interneuron} AVBL & \cellcolor{interneuron} AVBR & \cellcolor{interneuron} AVDL & \cellcolor{interneuron} AVDR & \cellcolor{interneuron} AVEL & \cellcolor{interneuron} AVER & \cellcolor{interneuron} AVJL & \cellcolor{interneuron} AVJR & \cellcolor{interneuron} BDUL & \cellcolor{interneuron} BDUR & \cellcolor{interneuron} PVCL & \cellcolor{interneuron} PVCR & \cellcolor{interneuron} PVPL & \cellcolor{interneuron} PVPR \\
    \hline
    \cellcolor{interneuron} PVR & \cellcolor{interneuron} PVT & \cellcolor{interneuron} RIAL & \cellcolor{interneuron} RIAR & \cellcolor{interneuron} RIBL & \cellcolor{interneuron} RIBR & \cellcolor{interneuron} RIFL & \cellcolor{interneuron} RIFR & \cellcolor{interneuron} RIGL & \cellcolor{interneuron} RIGR & \cellcolor{interneuron} RIH & \cellcolor{interneuron} RIML & \cellcolor{interneuron} RIMR & \cellcolor{interneuron} RIPL \\
    \hline
    \cellcolor{interneuron} RIPR & \cellcolor{interneuron} RIR & \cellcolor{modulatory} ADEL & \cellcolor{modulatory} ADER & \cellcolor{modulatory} AIML & \cellcolor{modulatory} AIMR & \cellcolor{modulatory} ALA & \cellcolor{modulatory} AVFL$^*$ & \cellcolor{modulatory} AVFR$^*$ & \cellcolor{modulatory} AVHL & \cellcolor{modulatory} AVHR & \cellcolor{modulatory} AVKL & \cellcolor{modulatory} AVKR & \cellcolor{modulatory} AVL$^*$ \\
    \hline
    \cellcolor{modulatory} CEPDL & \cellcolor{modulatory} CEPDR & \cellcolor{modulatory} CEPVL & \cellcolor{modulatory} CEPVR & \cellcolor{modulatory} DVC & \cellcolor{modulatory} HSNL$^*$ & \cellcolor{modulatory} HSNR$^*$ & \cellcolor{modulatory} PVNL$^*$ & \cellcolor{modulatory} PVNR$^*$ & \cellcolor{modulatory} PVQL & \cellcolor{modulatory} PVQR & \cellcolor{modulatory} RICL & \cellcolor{modulatory} RICR & \cellcolor{modulatory} RID \\
    \hline
    \cellcolor{modulatory} RIS & \cellcolor{modulatory} RMGL & \cellcolor{modulatory} RMGR & \cellcolor{motor} IL1DL & \cellcolor{motor} IL1DR & \cellcolor{motor} IL1L & \cellcolor{motor} IL1R & \cellcolor{motor} IL1VL & \cellcolor{motor} IL1VR & \cellcolor{motor} RIVL & \cellcolor{motor} RIVR & \cellcolor{motor} RMDDL & \cellcolor{motor} RMDDR & \cellcolor{motor} RMDL \\
    \hline
    \cellcolor{motor} RMDR & \cellcolor{motor} RMDVL & \cellcolor{motor} RMDVR & \cellcolor{motor} RMED & \cellcolor{motor} RMEL & \cellcolor{motor} RMER & \cellcolor{motor} RMEV & \cellcolor{motor} RMFL$^*$ & \cellcolor{motor} RMFR$^*$ & \cellcolor{motor} RMHL$^*$ & \cellcolor{motor} RMHR$^*$ & \cellcolor{motor} SIADL & \cellcolor{motor} SIADR & \cellcolor{motor} SIAVL \\
    \hline
    \cellcolor{motor} SIAVR & \cellcolor{motor} SIBDL & \cellcolor{motor} SIBDR & \cellcolor{motor} SIBVL & \cellcolor{motor} SIBVR & \cellcolor{motor} SMBDL & \cellcolor{motor} SMBDR & \cellcolor{motor} SMBVL & \cellcolor{motor} SMBVR & \cellcolor{motor} SMDDL & \cellcolor{motor} SMDDR & \cellcolor{motor} SMDVL & \cellcolor{motor} SMDVR & \cellcolor{motor} URADL \\
    \hline
    \cellcolor{motor} URADR & \cellcolor{motor} URAVL & \cellcolor{motor} URAVR & \cellcolor{sensory} ADFL & \cellcolor{sensory} ADFR & \cellcolor{sensory} ADLL & \cellcolor{sensory} ADLR & \cellcolor{sensory} AFDL & \cellcolor{sensory} AFDR & \cellcolor{sensory} ALML & \cellcolor{sensory} ALMR & \cellcolor{sensory} ALNL$^*$ & \cellcolor{sensory} ALNR$^*$ & \cellcolor{sensory} AQR$^*$ \\
    \hline
    \cellcolor{sensory} ASEL & \cellcolor{sensory} ASER & \cellcolor{sensory} ASGL & \cellcolor{sensory} ASGR & \cellcolor{sensory} ASHL & \cellcolor{sensory} ASHR & \cellcolor{sensory} ASIL & \cellcolor{sensory} ASIR & \cellcolor{sensory} ASJL & \cellcolor{sensory} ASJR & \cellcolor{sensory} ASKL & \cellcolor{sensory} ASKR & \cellcolor{sensory} AUAL & \cellcolor{sensory} AUAR \\
    \hline
    \cellcolor{sensory} AVM$^*$ & \cellcolor{sensory} AWAL & \cellcolor{sensory} AWAR & \cellcolor{sensory} AWBL & \cellcolor{sensory} AWBR & \cellcolor{sensory} AWCL & \cellcolor{sensory} AWCR & \cellcolor{sensory} BAGL & \cellcolor{sensory} BAGR & \cellcolor{sensory} DVA & \cellcolor{sensory} FLPL & \cellcolor{sensory} FLPR & \cellcolor{sensory} IL2DL & \cellcolor{sensory} IL2DR \\
    \hline
    \cellcolor{sensory} IL2L & \cellcolor{sensory} IL2R & \cellcolor{sensory} IL2VL & \cellcolor{sensory} IL2VR & \cellcolor{sensory} OLLL & \cellcolor{sensory} OLLR & \cellcolor{sensory} OLQDL & \cellcolor{sensory} OLQDR & \cellcolor{sensory} OLQVL & \cellcolor{sensory} OLQVR & \cellcolor{sensory} PLNL$^*$ & \cellcolor{sensory} PLNR$^*$ & \cellcolor{sensory} SAADL & \cellcolor{sensory} SAADR \\
    \hline
    \cellcolor{sensory} SAAVL & \cellcolor{sensory} SAAVR & \cellcolor{sensory} SDQL$^*$ & \cellcolor{sensory} SDQR$^*$ & \cellcolor{sensory} URBL & \cellcolor{sensory} URBR & \cellcolor{sensory} URXL & \cellcolor{sensory} URXR & \cellcolor{sensory} URYDL & \cellcolor{sensory} URYDR & \cellcolor{sensory} URYVL & \cellcolor{sensory} URYVR \\
    \hline
  \end{tabularx}
\end{table*}

\begin{figure}[htbp]
  \begin{minipage}{0.6\textwidth}
    \centering
    \includegraphics[width=9cm]{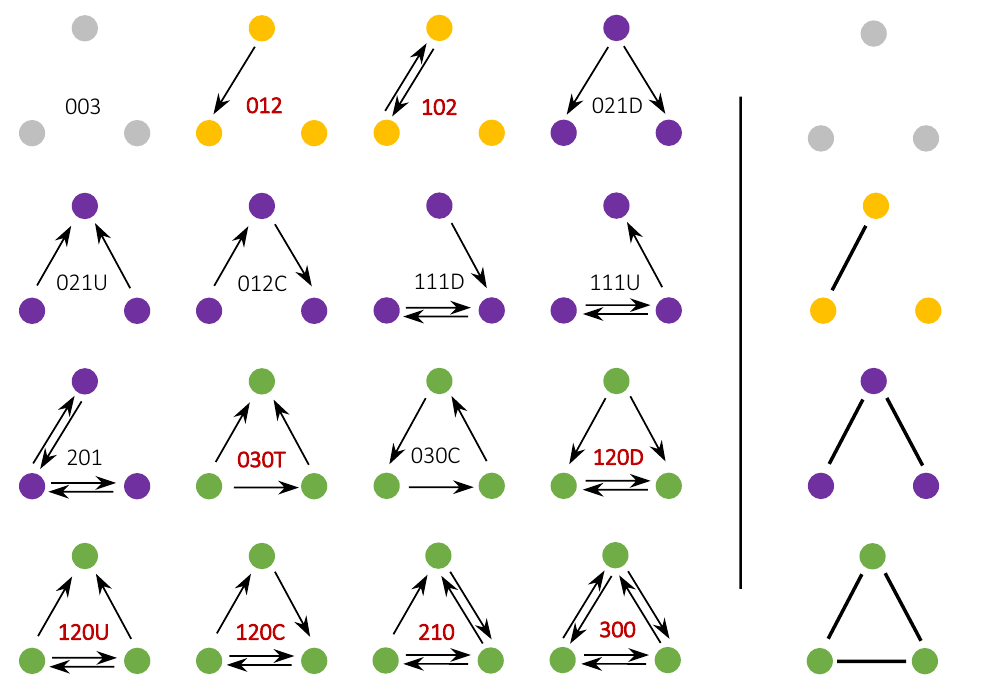}
  \end{minipage}%
  \hfill
  \begin{minipage}{0.4\textwidth}
    \caption{Triad census. \textbf{Left}: In a directed graph, there are $16$ distinctive connectivity patterns between three nodes. Each pattern is denoted by a three-digit code, with each digit representing the count of mutual links ($\leftrightarrow$), single links ($\rightarrow$), and non-existent links, respectively. Additionally, a letter may be appended to indicate if the pattern represents a cycle (C), a transitive (T), an upward (U), or a downward (D) connection. We have highlighted in bold red the codes of those patterns that are over-represented in the adult \emph{C.elegans} network (directed) of chemical synapses \cite{varshney2011,cook2019}. \textbf{Right}: In an undirected graph, $4$ unique connectivity patterns are possible among three nodes. We have employed a color-coding system to illustrate the correspondence between patterns when approximating a directed graph with an undirected one.}
    \label{f-triad_census}
    \vspace{1em}
  \end{minipage}
\end{figure}

% Directed -> Undirected, rationale
In MS, we used an ERG model, sec. \ref{s-ERGM}, to infer the parameters of an $F$-landscape, fig.(\ref{f-overview}). This required formulating the $F$ metric in terms of graph motifs. In doing this, a natural starting point is looking at the subnetwork distributions of simple motifs, as they are the building block of higher order structures. Previous graph analyses of the adult \ce\ network of synapses \cite{varshney2011,cook2019} studied the triad census, i.e., the counts of all directed connection patterns involving three nodes - there exist $16$ possible motifs, see fig.(\ref{f-triad_census}). These investigations found that the patterns involving the same number of empty dyads have similar statistical properties, i.e., they are under- or over-represented with respect to randomized null models. For instance, the triplets of nodes with connections between each pair (regardless of the directions) occur with greater frequency than it would be expected by chance - this is true for the motifs $030$T, $120$D, $120$U, $120$C, $210$, $300$ while $030$C  represents an exception. Based on this evidence, we simplified the analysis by considering the network of undirected synaptic connections of the \ce, fig.(\ref{f-triad_census}). As a coarse-graining procedure, this implies a loss of information. However, it allows to formulate an $F$ metric eq.(\ref{e-F-ce}) in terms of only a few sufficient statistics, avoiding the combinatorial proliferation of possible patterns for higher-order motifs in directed networks. This should be regarded as a first step towards more complex models of the \ce\ brain maturation, which will account for the directed nature of the information flow. 

% General features of the graphs
In tab.(\ref{t-sum_stats}), we provide a comprehensive summary of the properties exhibited by the eight undirected graphs analyzed in this study. We observe an increased neuronal connectivity over time, which is further reflected in the growth of the number of simple motifs such as $2$-paths and triangles. The rise in the clustering coefficient is consistent with the increasing formation and consolidation of functional circuits or modules \cite{witvliet2021}. It further suggests the deployment of a biological strategy for increasing the system's redundancy, thereby bolstering the resilience against potential failures. Moreover, we observe a notable increase in efficiency alongside a reduction in the shortest path lengths. This observation aligns well with the documented emergence of highly connected hubs \cite{witvliet2021}, leading to a more tightly woven neural network.

\begin{table*}[h!]
\centering
\caption{Properties of the \ce\ networks of undirected synaptic connections. Each row corresponds to a graph, the first (birth) and the last two (adulthood) are highlighted. We compute the number of \textbf{nodes}, \textbf{edges}, \textbf{two-star} -- or connected triples --, \textbf{triangles}, the average shortest path (\textbf{av.sh.path}) -- or average geodesic distance -- , the global efficiency (\textbf{glob.eff.}) and the average clustering coefficient (\textbf{clust.coeff.}). See also sec. \ref{ss-gof} for the definitions.} 
\label{t-sum_stats}
\begin{tabular}{c|cccccc|c}
\multicolumn{1}{c}{\textbf{t[h]}} & \multicolumn{1}{c}{\textbf{nodes}} & \multicolumn{1}{c}{\textbf{edges}} & \multicolumn{1}{c}{\textbf{two-star}} & \multicolumn{1}{c}{\textbf{triangles}} & \multicolumn{1}{c}{\textbf{av.sh.path}} & \multicolumn{1}{c}{\textbf{glob.eff.}} & \multicolumn{1}{c}{\textbf{clust.coeff.}} \\ \midrule
\rowcolor[HTML]{F2EAD3}
$\bm{0}$ & $161$ & $617$ & $5976$ & $346$ & $2.993$ & $0.380$ & $0.208$ \\
$\bm{5}$ & $162$ & $782$ & $9273$ & $601$ & $2.712$ & $0.416$ & $0.232$ \\
$\bm{8}$ & $162$ & $788$ & $9299$ & $614$ & $2.712$ & $0.416$ & $0.245$ \\
$\bm{16}$ & $168$ & $907$ & $11838$ & $830$ & $2.617$ & $0.428$ & $0.246$ \\
$\bm{23}$ & $173$ & $1166$ & $18449$ & $1406$ & $2.430$ & $0.459$ & $0.262$ \\
$\bm{27}$ & $174$ & $1175$ & $18866$ & $1433$ & $2.429$ & $0.458$ & $0.274$ \\
\rowcolor[HTML]{DFD7BF}
$\bm{45}$ & $180$ & $1633$ & $34124$ & $2889$ & $2.217$ & $0.498$ & $0.286$ \\
\rowcolor[HTML]{DFD7BF}
$\bm{45}$ & $180$ & $1669$ & $35677$ & $3003$ & $2.206$ & $0.501$ & $0.292$ \\
\bottomrule
\end{tabular}
\end{table*}

\subsection{The EE dynamics as brain wiring dynamics: an interpretation} \label{ss-interpr}

In this section, we illustrate how the EE dynamics eq.(\ref{e-mut}-\ref{e-sel}) could be interpreted from the point of view of the individual developing system. In so doing, we aim to show how the EE dynamic is consistent with the view of brain wiring dynamics proposed in \cite{hassan2015}, as essentially driven by a set of (genetically encoded) rules that allow neurons to make local decisions about whom to connect with. Note that we take the network representation of the brain as correct and focus on its growth (edge addition) during development. Therefore, a model of the brain wiring process at the molecular level (axon guidance, growth factors, etc.) is neither possible nor desired here. % B.1

\begin{figure}[htbp]
\centering
\includegraphics[width=15.5cm]{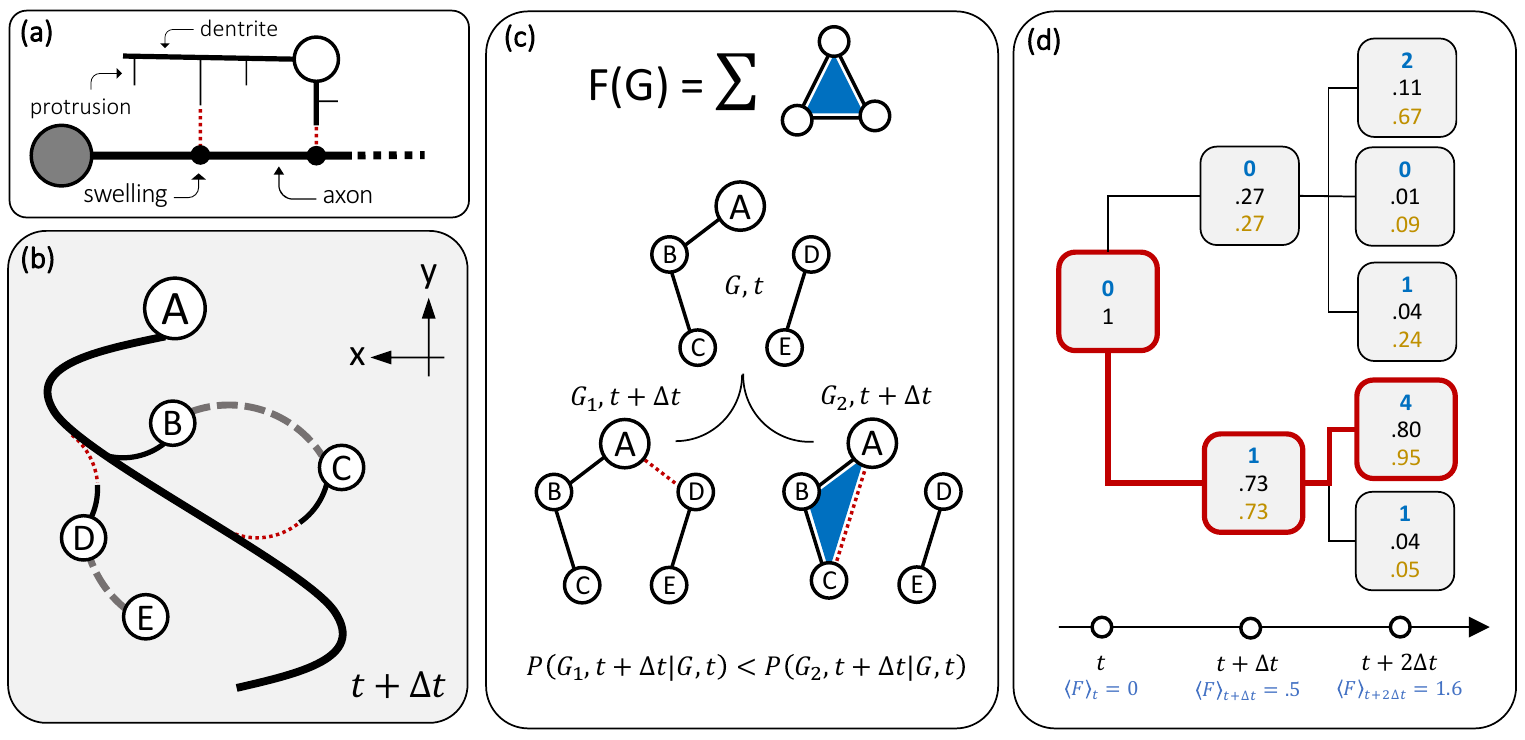}
\caption{Individual-based interpretation of the EE dynamic. 
(a) Establishment of a synapse, schematics. Presynaptic sites, presented as en passant swellings (black circles), appear on the axon shaft (thick black line) of a presynaptic neuron (gray circle). Postsynaptic neuronal processes - namely, dendrites (black lines) and their spine-like protrusions (thin black lines) - sprout from a postsynaptic neuron (white circle). Occasionally, they establish synaptic connections with physically proximal presynaptic sites (red dashed lines). The presynaptic processes of the postsynaptic neuron and the postsynaptic processes of the presynaptic neuron have not been depicted. 
(b) A simple scenario. One presynaptic neuron $A$ and four postsynaptic neurons $B,C,D,E$ are represented. The thick black line represent the axon elongating from $A$. At time $t$ a synaptic connection exists between the nodes $AB$ (black line connecting $B$ to the axon shaft). Additional connections exist between the neurons $BC$ and $DE$ (summarized by gray dashed line). After a time interval $\Delta t$, postsynaptic neuronal processes extend from the neurons $C,D$ towards the axon, that might develop into new connections (red dotted lines). On the contrary, neuron $E$ does not exhibit the emergence of any such postsynaptic process. 
(c) Graph representation of the scenario in (b). We hypothesise an $F$ metric that simply counts the number of triangles in the undirected graph representation of the neural system. The two potential connections between $AD$ and $AC$ at time $t+\Delta t$ can be represented as two different graph configurations, $G_1, G_2$, associated to different $F$ values. $G_2$, by virtue of its higher $F$, will be observed with higher probability.
(d) Example of two time steps of the EE dynamic ($\Delta t =1$ for simplicity). Each square represents a graph. In blue, we indicate the $F$ values. In black, the unconditioned probabilities computed at each time as $\exp [F(G_i)]/\sum_j \exp [F(G_j)]$ where the sum runs over all graphs at that time (column). In brown, the probabilities conditioned on the previous time-point. They can be computed either as above, restraining the sum to those graphs that come from the same parent graph at the previous time, as done in eq.(\ref{e-expl_ce}), or starting from the unconditioned probabilities and using $P(G_i,t+1|G_j,t)=P(G_i,t+1\cap G_j,t)/P(G_j,t)$, where $P(G_j,t)= \sum_k P(G_k,t+1\cap G_j,t)$. In bold-red we highlight the most likely developmental pattern.
}
\label{f-interpretation}
\end{figure}

The biological process we consider in this work - i.e., the formation of neuronal circuits - entails a multitude of complex molecular and cellular events, not fully understood \cite{colon2009}. In the case of the \ce, synapses are mainly found between physically proximal neuronal processes - the latter term generally refers to any projection from the cell body: dendrites, postsynaptic or axons, presynaptic. More in detail, most of the neurons have a simple morphology, their processes (one or two per each neuron) run in parallel bundles along the worm's body \cite{white1986}. Across development, they extend under the influence of molecular guidance cues. Presynaptic sites appear as \emph{en passant} swellings along the axon shaft. Postsynaptic neuronal processes are either dendrites or spine-like protrusions that grow out from dendrites \cite{witvliet2021}. Occasionally, dendrites or protrusions establish new synaptic connections, fig.(\ref{f-interpretation}a).

To demonstrate how the synapse formation process could be encompassed by the EE dynamics, let us take into account the simple scenario illustrated in fig.(\ref{f-interpretation}b). Let us suppose that, at a given time $t$, synaptic connections exist between neurons $AB$, $DE$, and $BC$. In the time interval $\Delta t$, postsynaptic neuronal processes from both neurons $C$ and $D$ emerge (or extend sufficiently), potentially leading to the formation of new synaptic connections with $A$. Contrarily, no such process appears on the neuron $E$, which therefore has no chance of establishing a connection with $A$. Provided that $\Delta t$ is sufficiently small, only one of the two possible synaptic connections $AC$ or $AD$ is likely to be observed. Although the choice is inherently stochastic, a connection offering a higher functional advantage will be associated to a higher probability. Under the hypothesis that the notion of \emph{biological function} is represented by the number of triangles in the undirected graph representation of the system, fig.(\ref{f-interpretation}c), we anticipate the preferential formation of the $AC$ connection, as it implies the formation of the $ABC$ triangle. In this sense, the "less functional" mutation $AD$ is penalised.

In this interpretation, a mutation event does not correspond to the establishment of a \emph{physical} connection, but to its precondition (formation/elongation of neuronal process), therefore to a \emph{potential} connection -- note that this marks a difference with respect to the evolutionary dynamics where different genetic mutations correspond to different physical individuals. The exploitation on the other hand acts on configurations that have been allowed by mutations. If $G$ is the graph configuration at time $t$ and $\bm{\tilde{G}}$ are the potential configurations allowed by mutations at time $t+\Delta t$, then
    \begin{equation}\label{e-expl_ce}
        P(G_i,t+\Delta t|G,t) = e^{\Delta t F(G_i)}/\sum_{G_j\in\bm{\tilde{G}}} e^{\Delta t F(G_j)} \ ,
    \end{equation}
where we have taken $\varphi=1$ for simplicity. Note that the above probabilities are conditioned on the configuration $G$ observed at time $t$.

The one above illustrated is an EE dynamic of a single, developing system. From the standpoint of the individual neuron, the process of synaptogenesis consists of a series of stochastic decisions about which other neuron to connect with. These decisions are biased towards those connections that lead to higher functional gains, which in turn are evaluated based on the information available to the neuron at any given time. Note that for this interpretation to be possible, it is essential to define the biological function $F$ in terms of local network structures -- e.g., the triangles in the previous example -- because neurons only have knowledge of their neighbourhood, not of global system properties. 

The EE dynamic contains a comprehensive information about the possible dynamical trajectories, being defined for the probability distribution rather than for a single evolving system. In particular, it allows to compute the (unconditioned) probability of all possible configuration that might have appeared at time $t$, including those that might have taken very unlikely developmental paths. The difference is illustrated in fig.(\ref{f-interpretation}d). This provides a justification for employing an EE dynamics to encapsulate the inter-individual variations observed in brain wiring. \\

\section{\ce\ model fit}

The model of the \ce\ brain growth illustrated in \ms\ has only six parameters, four describe the $F$-landscape, two the EE dynamics. Their values must be deduced from the data, as detailed in sec. \ref{ss-ergm_fit}-\ref{ss-dev_traj}. Only the birth and adult snapshots of the \ce\ brain are employed for this purpose. The assessment of how well a model's predicted outcomes aligns with the actual observed data is discussed in sec. \ref{ss-gof}. 

Let us also emphasize the philosophy behind our inferential scheme. According to our argument, both the topography of the functional landscape and the parameters of the dynamics are ultimately genetically encoded. This aligns with the existence of genetically encoded developmental algorithms that drive the brain growth \cite{hassan2015}. All the system "knows" about the developmental process is this algorithm. The evolutionary history of the species has selected such an algorithm so that an adult, functional brain can be achieved at the end of the developmental process. This is a biological fact. To enforce it in our model, we make the natural choice of learning and constraining the properties of the developmental algorithm by using the endpoint of the biological process (adult stage). We have no arguments to exclude a priori alternative choices -- e.g. learning the features of the developmental algorithm by using only the birth stage, or postulating them a priori or more. In the latter cases, however, it remains an open question how to obtain the correct functional brains in adulthood. In \ms, we have shown that our choice not only yields adult functional brains (by construction), but also leads to an effective reconstruction of the entire developmental trajectory.

\subsection{\emph{F}-landscape}\label{ss-ergm_fit}
The inference of the $F$-landscape was carried out through exponential random graph (ERG) models, sec. \ref{s-ERGM}, using the two adult snapshots of the \ce\ network of chemical synapses. The ERG inference is defined for a single graph, let us call $G^*_{T,i}$ a generic adult snapshot. The model described in eq.(\ref{e-stats}, \ref{e-F-ce}) can be summarized in the following ERG Hamiltonian $\mathcal{H}$:
\begin{equation}\label{e-H-ERGM}
    - \mathcal{H}(G^*_{T,i}) = \theta_{gwd}\ e^{\tau_d}\sum_{k=1}^{N-1} \Big\{1-\big(1-e^{-\tau_d}\big)^{k}\Big\}x_{d}^{(k)}(G^*_{T,i}) + \theta_{gwesp} \ e^{\tau_{esp}}\sum_{k=1}^{N-2} \Big\{1-\big(1-e^{-\tau_{esp}}\big)^{k}\Big\}x_{esp}^{(k)}(G^*_{T,i})\ ,
\end{equation}
where we have used the explicit formulation of the statistics \texttt{gwd} eq.(\ref{eTERM-GWD}) and \texttt{gwesp} eq.(\ref{eTERM-GWESP}). The above expression has four parameters, namely $\theta_{gwd},\theta_{gwesp}$ and $\tau_d,\tau_{esp}$. Note that from a theoretical point of view, they are fundamentally different. The former are the linear weights of the Hamiltonian that originate from the maxent origin of the ERG probability distribution, sec. \ref{s-ERGM}. Conversely, the latter dictate the formal definition of the graph statistics, therefore they can be ascribed to the problem of model selection. In practice, however, they can be estimated simultaneously by the \texttt{ergm} package \cite{krivitsky2022}. We used the following specification:

\begin{lstlisting}[caption={Specifying an ERG model based on the Hamiltonian eq.(\ref{e-H-ERGM}) using the \texttt{ergm} package. \texttt{G} is the graph to be used for inference, \texttt{fixed=F} implies that the decay parameters of the curved statistics are to be estimated. The model is constrained to graphs that have the same number of edges as the \texttt{G}. As initial guess of the four parameters to be estimated, we provide \texttt{(1,1,1,1)}. For the estimation, we have used \texttt{Rstudio v2022.12.0.353}, \texttt{R v4.0.4} and \texttt{ergm v4.3.2}. The scripts ara available in the Github folder \cite{dichio2023git}.}]
# ERGM formula
fit <- ergm(formula = G ~ gwdegree(fixed=F)+gwesp(fixed=F), 
            constraints = ~ edges,
            control=snctrl(init = c(1,1,1,1))
            )
\end{lstlisting}

Note that in the specification of the formula, we constrained the space of graphs explored by the \texttt{ergm} numerical routines to those graphs that have the same number of nodes as the input graph. The reason for this is that the number of edges of the graphs in the EE dynamics is controlled by the exploration rate and therefore it is not a degree of freedom at the disposal of the ERG inference. The results of the estimations are summarized in tab.(\ref{t-ERGM-estimation}).
\begin{table}[h]
\renewcommand{\arraystretch}{1.5}
\setlength{\arrayrulewidth}{2pt}
\centering
\begin{tabular}{c>{\columncolor[HTML]{F2EAD3}}cc>{\columncolor[HTML]{F2EAD3}}cc}
& $\theta^*_{gwd}$ & $\lambda^*_{gwd}$ & $\theta^*_{gwesp}$ & $\lambda^*_{gwesp}$\\
\hline
$G^*_{T,1}$ &  $0.45 \pm 0.20$ & $1.91 \pm 0.46$ &  $0.626 \pm 0.056$ & $1.432 \pm 0.067$ \\
\hline
$G^*_{T,2}$ &  $0.43 \pm 0.20$ & $1.97 \pm 0.48$ & $0.529 \pm 0.048$ &  $1.542 \pm 0.075$\\
\hline
$\overline{\text{ERG}}$ &  $\bm{0.44} \pm 0.14$ & $\bm{1.94} \pm0.33$ &  $\bm{0.578} \pm 0.037$ & $\bm{1.487} \pm 0.050$\\\bottomrule
\end{tabular}

\caption{Estimated parameters of the ERG model eq.(\ref{e-H-ERGM}) from the two adult \ce\ (undirected) networks of chemical synapses. The maxent parameters $\bm{\theta}^*$ are both significant and positive for all networks. The parameters $\bm{\lambda}^*$ controlling for the geometric decays of the model statistics are significant -- and positive by construction. The third row represents the mean-ERG, final parameter estimations for the EE dynamic are boldfaced.}
\label{t-ERGM-estimation}
\end{table}

An output procedure is needed to combine the estimations from the group of adult individuals in our dataset. Considered that each individual yields independent estimation of the ERG parameters, a simple choice is to take the average estimations, i.e., 
\begin{equation}\label{e-erg_estimations}
    \theta_{gwd}^* = 0.44,\ \tau_d^* = 1.94,\ \theta_{gwesp}^* = 0.578,\ \tau_{esp}^* = 1.487\ 
\end{equation}
In the ERG  literature, this procedure is referred as mean-ERG \cite{dichio2022}. Here, we are essentially limited by the availability of only two individual connectomes.

\subsection{Developmental trajectory}\label{ss-dev_traj}

The theoretical picture of the \ce\ brain wiring dynamics discussed in \ms\ is that of a stochastic, state-dependent dynamics in the functional landscape $F(G)$ as in eq.\eqref{e-F-ce}, with parameters eq.\eqref{e-erg_estimations}.
The long-term dynamics of this model is quite simple: if we waited long enough, we would end up with (almost) full graphs, independently of our starting point. However, we do not wait long enough. Instead, the edge formation is controlled by the exploration rate $\mu$, and set to reach the appropriate number of edges at adulthood. Therefore, although the maximisation of the $F$ metric learned in the adult stage drives the dynamics, the adult stage itself is not intended to be a maximum of the $F$ landscape. 

In \ms, we have already discussed the choice of the boundary conditions and the inference of the exploration rate $\mu^*$ (growth only). This leaves us with only one degree of freedom, the functional pressure $\rho$, which we discuss in sec. \ref{ss-dev_traj}.2. Running a single simulation of the developmental trajectory of the \ce\ requires a few additional specifications. 

\subsubsection{A single run}\label{sss-a_run}

If the theory is completely specified by the above parameters, the simulations described in sec. \ref{s-sim} have two additional degrees of freedom.

The first one, $M$ is the number of samples drawn from the graph probability distribution at each simulation step. Small number of samples will result in high level of noise for the statistics computed as in eq.(\ref{e-emp-av}). Conversely, a tangible constraint arises from the computational load, which scales linearly with $M$. In the simulations described in this section, we employed $M=1024$ and empirically verified that the outcomes are not significantly altered when simulations are run with $M=2048, M=4096$. The second internal degree of freedom is the inverse time-step $\nu=\Delta t^{-1}$. We fixed it by requiring that in one simulation step a single edge addition was to be observed in each graph within our population. This means choosing $\Delta t = (L\mu^*)^{-1}\sim 4.34\times10^{-2}\ h $, hence $\nu \sim 23$. Once again, the computational time scales linearly with $\nu$. 

Our EE simulations currently do not support a node dynamic (appearance/disappearance). Consequently, to set the initial conditions, we embedded the birth connectome, consisting of $161$ nodes, within a larger network that matched the adult connectome's node count of $180$.

A single simulation with the configuration here presented takes approximately $1.5$ hours. The scripts are available in the Github folder \cite{dichio2023git}.

\subsubsection{Optimal functional pressure}
In \ms, we proposed to fix the degree of freedom of the functional pressure $\rho$ by requiring that our simulations optimally reproduce the experimental graph statistics at the adult age. The EE simulated dynamics allows to compute at each time $t$ a whole distribution of graph statistics, while experimental data consists of one or two isolated point per each $t$. A natural generalization of the Euclidean distance to the distance between a multivariate distribution $Q$ on $\mathbb{R}^r$ and point $\bm{y}^*\in\mathbb{R}^r$ is the Mahalanobis distance $\delta^{mah}_Q$ \cite{mahalanobis2018}, defined as
\begin{equation}\label{e-mah-def}
    \delta^{mah}_Q = \sqrt{ (\av{\bm{y}}_Q-\bm{y^*})^{\top}\Sigma^{-1}_Q (\av{\bm{y}}_Q-\bm{y^*}) }
\end{equation}
where $\av{\bm{y}}_Q$ and $\Sigma_Q$ are the mean and covariance matrix of $Q$. $\delta^{mah}_T$ used in MS to find the optimal $\rho^*$ corresponds to eq.(\ref{e-mah-def}) where $Q$ is the two-dimensional distribution of graph statistics eq.(\ref{e-stats}) at the adult age $T = 45\ h$ (we use the time $T$ to label the distribution), $\av{\bm{y}}_Q = \av{\bm{x}}_T$ and $\Sigma_Q = \Sigma_T$ are the ensemble average and covariance matrix, while $\bm{y}^*=\bar{\bm{x}}(\bm{G}_T^*)$ are the experimental graph statistics, averaged over the two adult connectomes, i.e., $y_i^* = [ x_i(G^*_{T,1}) + x_i(G^*_{T,2})]/2$. 

\begin{figure}
    \centering
    \subfloat[\label{f-statspace}]{\includegraphics[height=8.6cm]{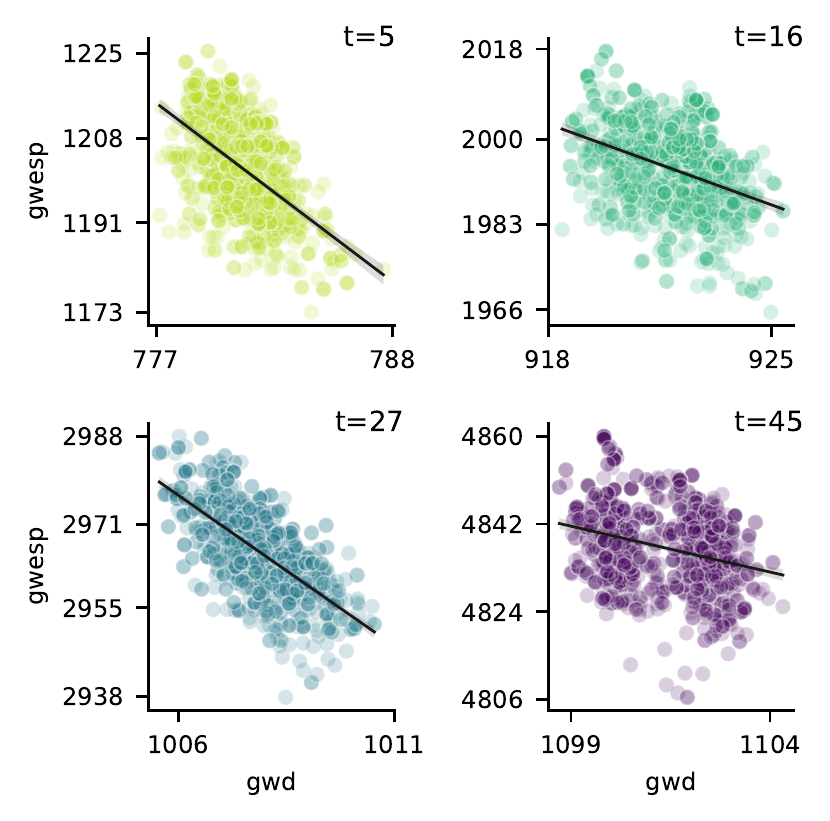}
    }\hfill
    \subfloat[\label{f-mah-allt}]{\includegraphics[height=8.cm]{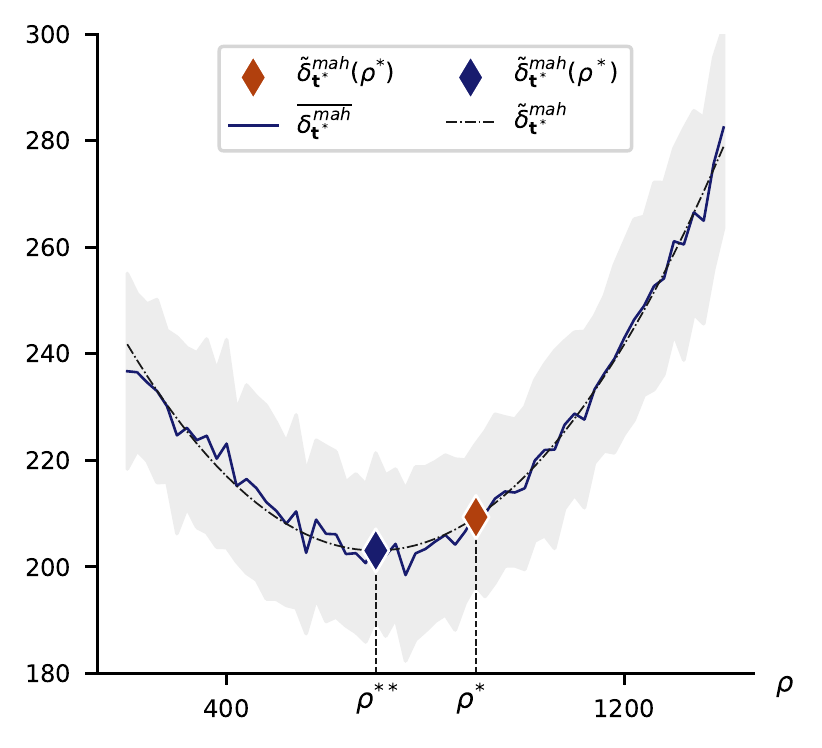}
    }
  \caption{(a) Population-based simulations, in the space of model statistics eq.(\ref{e-stats}). At any given time $t$ our simulation allow to extract a two-dimensional distribution in the space \texttt{gwd-gwesp}.  We show those corresponding to four of the seven experimental time-stamps. Note that the statistics are (anti-)correlated, in computing a distance from the distribution, the covariance structure must be taken into account. Here we used the same simulation as in fig.(\ref{f-results}b-c) -- these are the underlying distributions of the average values plotted in fig.(\ref{f-results}b). (b) We run $100$ simulations $\forall \rho\in \{200+20\ i, 0\le i\le60 \}$. For each $\rho$, we compute the mean and standard deviation of $\delta_{\bm{t}^*}^{mah}$ (blue line and shaded area, respectively). The data are fitted with a parabolic curve $\tilde{\delta}_{\bm{t}^*}^{mah}$ ($R^2=.99$) and its minimum $\rho^{**}$ is highlighted (blue diamond). On the same curve, we show the position of $\tilde{\delta}_{\bm{t}^*}^{mah}(\rho^{*})$, the value of the all-time Mahalanobis distance that corresponds to the functional pressure $\rho^*$ we selected in \ms, based on the adult stage exclusively (orange diamond). The two overlap within the error bars. This plot is complementary to fig.(\ref{f-results}a) in \ms.
  }\label{f-mah-dist}
\end{figure}

The Mahalanobis distance, as defined in eq.(\ref{e-mah-def}), has two intriguing characteristics that render it particularly appropriate for the task at hand. 
\begin{itemize}
    \item[(i)] It takes into account the covariance structure of the multivariate distribution. This is crucial,  since the statistics eq.(\ref{e-stats}) are clearly not independent, fig.(\ref{f-statspace}). In particular, let us consider the \emph{whitening} transformation \cite{kessy2018}  
    \begin{equation}\label{e-whitening}
        \bm{y}\rightarrow \bm{z}=\Sigma_Q^{-\frac{1}{2}}\bm{y}\qquad   \forall\bm{y}\in\mathbb{R}^r
    \end{equation}
     where $\Sigma_Q^{-\frac{1}{2}}$ is the inverse principal square root of the covariance matrix $\Sigma_Q$. It is easy to show that (a) the transformed variables have unit diagonal (white) covariance matrix and (b) eq.(\ref{e-mah-def}) corresponds to the Euclidean distance of the transformed variables. In words, the Mahalanobis distance accounts for covariance structure by computing the Euclidean distance of the \emph{whitened} (standardized) data. 
    \item[(ii)] It is scale invariant, i.e., it is invariant under affine transformations $\bm{y}\rightarrow A \bm{y} + \bm{b} \ \ \forall \bm{y}\in\mathbb{R}^r$, where $A$ is an $r\times r$ matrix and $\bm{b}\in\mathbb{R}^r$. For instance, it would not change if we scaled the statistics by their corresponding ERG model parameter. Therefore, eq.(\ref{e-mah-def}) provides a common ground for comparing models defined by different sets of $\bm{\theta}$ -- note that the same is not be true if instead we consider a distance function based on the $F$ metric.
\end{itemize}

In order determine an optimal functional pressure, we started by scanning uniformly possible values of $\rho$ -- $100$ runs for each. We computed $\delta_T^{mah}$ for each simulation, averages and standard deviations are depicted in fig.(\ref{f-results}a). As expected, for mild functional pressures, the simulations do not attain the desired value at adulthood, hence the Mahalanobis distance increases for low values of $\rho$. Furthermore, it also increases for too high values of $\rho$, implying an overestimation of the functional constraints deployed by the biological system. 

We fitted the data with a quadratic curve $\Tilde{\delta}_T^{mah}(\rho) = a\rho^2+b\rho+c$ and took the position of the minimum $-b/2a$ as optimal functional pressure $\rho^*$. This last methodological step brought the inference process to completion. Finally, in order to showcase the results of a single simulation of the \ce\ brain maturation in fig.(\ref{f-results}c-\ref{f-results}d), we ran $500$ simulations with the inferred parameters $\mu^*,\rho^*$ and again selected as best simulation the one that minimized $\delta_T^{mah}$.

Similarly to eq.(\ref{e-mah-def}), one can compute the distance $\delta^{mah}_t$ of the simulated distribution of statistics from the corresponding experimental values at all observed time points $\bm{t}^*=\SIlist{5;8;16;23;27;45}{\hour}$. Note that by assumption $\delta^{mah}_{t=0}=0$ and that, except for the adult age, only one experimental graph is available, therefore $\bm{y}^*=\bm{x}(G^*_t)$. The total error over the whole (observed) time series can be simply defined by
\begin{equation}\label{e-mah-allt}
    \delta^{mah}_{\bm{t}^*} = \sum_{t \in \bm{t}^*, \ t>0 } \delta^{mah}_t \ .
\end{equation}
Using the latter, we repeated the same steps described in the previous paragraph. The minimum of the quadratic fit  of $\Tilde{\delta}^{mah}_{\bm{t^*}}(\rho)$ was found to be located at $\rho^{**}=7.001\times10^2$. In fig.(\ref{f-results}a), we plotted  $\Tilde{\delta}_T^{mah}(\rho^{**})$, i.e., the average value of the Mahalanobis distance defined using the adult age exclusively that we would have obtained if we had chosen $\rho^{**}$. This value fell within one error bar from $\Tilde{\delta}_T^{mah}(\rho^{*})$. In fig.(\ref{f-mah-allt}), we corroborate this result by showing the plot complementary to fig.(\ref{f-results}a), for eq.(\ref{e-mah-allt}).

\subsection{Feature generalization}\label{ss-gof}
By feature generalization, we refer to the ability of a model to reproduce other features of the data than those included in the model formulation. We here briefly provide definitions of the metrics used in this work to this purpose, see also \cite{latora2017} for a general reference. The package \texttt{NetworkX v2.6.3} with \texttt{Python 3.9.7} has been used, the scripts are available in the Github folder \cite{dichio2023git}. 

Let us consider an unweighted, undirected graph $G$ with $N$ nodes, and adjacency matrix $A = \{a_{ij}\}$. Let $k_i = \sum_j a_{ij}$ be the degree of the the $i$-th node.

\begin{figure}
    \centering
    \subfloat[\label{f-gof_stats}]{\includegraphics[height=7.24cm]{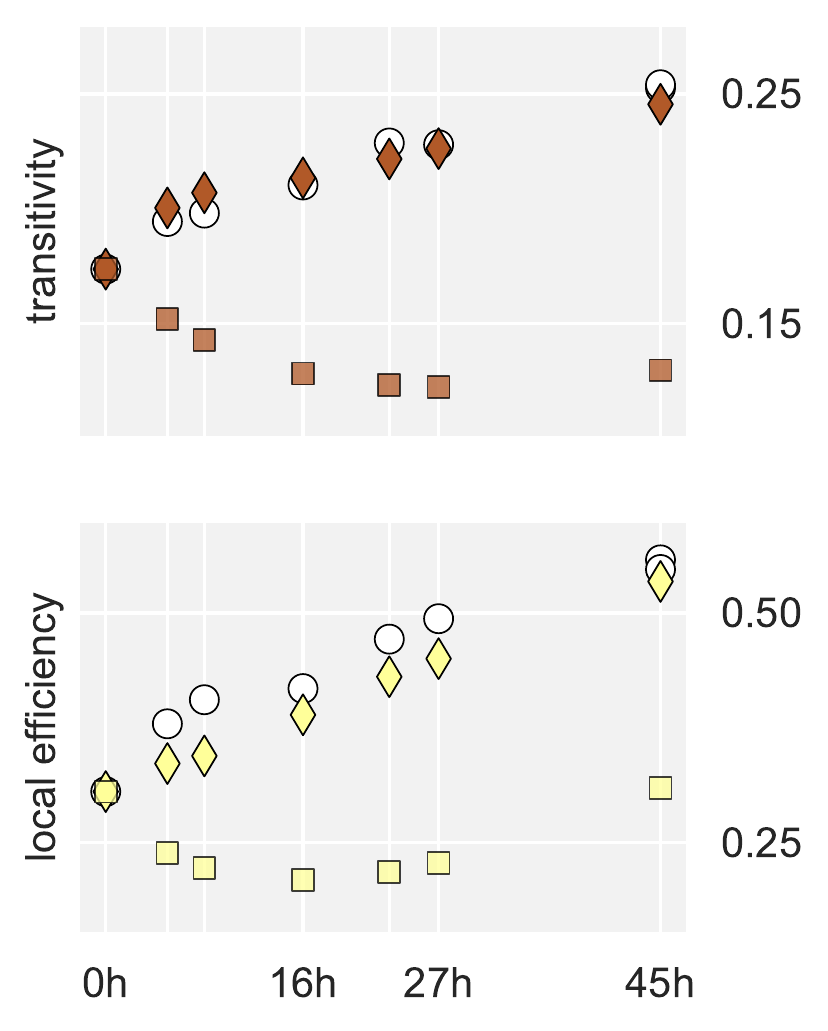}
    }
    \subfloat[\label{f-gof_cumdeg}]{\includegraphics[height=7.7cm]{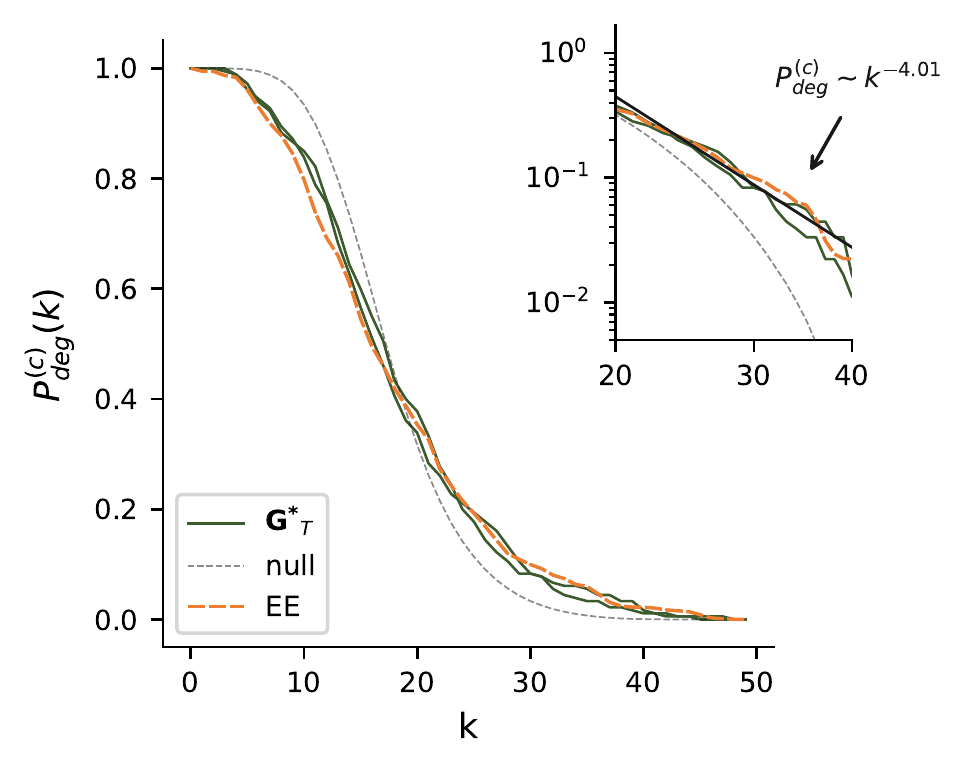}
    }
  \caption{Feature generalization assessment of the EE dynamics. (a) Transitivity (brown) and local efficiency (yellow). Given a statistic $y$, at each experimental time point $t\in\bm{t}^*$ we compute 
  its experimental value $y(G^*_t)$ (circles), the average $\av{y(G)}_t$ (diamonds) based on our population-based simulations eq.(\ref{e-emp-av}), and the same average with simulations based on a null growth model with $\rho=0$ (squares). In both cases our simulations closely track the experimental values. (b) Cumulative degree distributions at adulthood, $T = \SI{45}{\hour}$. Those corresponding to the two adult \ce\ brain graphs are represented as green solid lines. Dashed orange line for the simulated cumulative degree distribution, bin-wise average over the distribution of graphs we obtain from simulations. The Kolmogorov-Smirnov statistics between the latter and the two empirical distributions are $D_{KS} = 0.08,\ 0.10$, respectively. Gray dashed line for the corresponding values from a null model with $\rho=0$, $D_{KS} = 0.24,\ 0.18$. Inset: zoom in the high-degree tail, loglog plot. The black line is a linear fit of the simulated high-degree tail of the distribution.
  }\label{f-gof}
\end{figure}

\begin{itemize}
    \item \emph{Clustering}. A first family of graph metrics aims at computing the extent to which nodes in a graph tend to cluster together. The most straightforward manifestation of clustering is a higher-than-random probability that two nodes that are connected to a common node are also connected to each other. Such a behaviour can be quantified by the following two metrics:
    \begin{itemize}
    \item[$\circ$] \emph{Transitivity}, fig.(\ref{f-gof_stats}). It evaluates the ratio between the number of existing triangles -- triples $(i,j,k)$ with $a_{ij}=a_{jk}=a_{ik}=1$ -- and the number of connected triples, i.e., the triples $(i,j,k)$ with $a_{ij}=a_{jk}=1$. The symmetry factor accounts for the fact that each triangles has $3$ connected triples. Formally,
        \begin{equation}\label{e-transitivity}
            T = \frac{3\times \text{\# triangles} (G)}{\text{\# connected triples}(G)} = \frac{tr(A^3)}{\sum_{i\ne j}(A^2)_{ij}}\ .
        \end{equation}
    \item[$\circ$] \emph{Average clustering coefficient}, fig.(\ref{f-results}d). It is defined as:
    \begin{equation}\label{e-clustcoeff}
        C = \frac{1}{N} \sum_{i=1}^N C_i\ , \qquad C_i = \frac{\text{\# connected pairs of neighbors of } i}{\text{\# pairs of neighbors of } i} = \frac{\sum_{j,l}a_{ij}a_{jl}a_{il}}{k_i(k_i-1)}\ .
    \end{equation}
    where $k_i=\sum_ja_{ij}$ is the degree of the node $i$. In words, $C$ is the average of the local clustering coefficient $C_i$. The latter is computed by considering the subnetwork induced by the node $i$ and its first neighbors and quantifies the relative number of neighbors of $i$ that are also themselves neighbors. 
\end{itemize}
    Albeit both $T$ and $C$ take values in the interval $[0,1]$ and reach $1$ in the case of perfect transitivity, they are not equivalent. The clustering coefficient is more influenced by low-degree nodes (since it averages over all nodes), while transitivity is more influenced by high-degree nodes (since it considers all possible triangles in the network). Therefore, transitivity might be more representative of the overall network structure, while the clustering coefficient could provide more insight into local structures or subnetworks within the network.

    \item \emph{Efficiency.} A second family of metrics aims at quantifying the effectiveness of information or resource exchange over the network. As a general idea, the closer are two nodes in the graph, the more efficiently information will be exchanged among them.
    \begin{itemize}
        \item[$\circ$] \emph{Global efficiency}, fig.(\ref{f-results}d). A convenient way to quantify a graph's efficiency is to compute the harmonic mean of geodesic lengths. More in detail, let $D$ be the matrix whose elements $d_{ij}$ represent the shortest path (geodesic) from the node $i$ to $j$ -- by definition; $d_{ij}=\infty$ for disconnected nodes. The \emph{global efficiency} $E_{g}$ is defined as 
        \begin{equation}\label{e-efficiency}
            E_{g} = \frac{1}{N(N-1)}\sum_{i \ne j}\frac{1}{d_{ij}}\ .
        \end{equation}
        A network with high global efficiency is typically characterized by short paths between any given pair of nodes, meaning information or resources can be disseminated rapidly across the network. This property is often observed in random or in 'small-world' networks. 
        \item[$\circ$] \emph{Local efficiency}, fig.(\ref{f-gof}a) . A complementary definition of efficiency can be given by averaging over all nodes a local notion of efficiency. The \emph{local efficiency} $E_l$  can be defined as 
        \begin{equation}
            E_l = \frac{1}{N} \sum_i E_g^{(i)}\ , 
        \end{equation}
        where $E_g^{(i)}$ is the global efficiency of the subgraph induced by the node $i$ and its neighbors. By looking at the efficiency of each node's immediate subnetwork, local efficiency provides insight into the network's robustness or resilience to failures or attacks. If a network has high local efficiency, the removal of a node would not significantly disrupt communication between its neighboring nodes. 
    \end{itemize}
    \item \emph{Node degrees}. The degree distribution of a network is a critical aspect to consider when studying network structures, as it encapsulates fundamental information about the network's structure, robustness to failures, and information spreading dynamics, fig.(\ref{f-gof_cumdeg}).

    Given a network, one may look directly at its degree distribution $P_{deg}(k) = n_k/N$ where $n_k$ is the number of nodes   with degree $k$, or its cumulative distribution 
        \begin{equation}
            P_{deg}^{(c)}(k) = P_{deg}(j\ge k) = \frac{1}{N}\sum_{j\ge k}n_j\ , \qquad n_j = \text{\# nodes with degree } j\ .
        \end{equation}
    Cumulative degree distributions are frequently employed in network analysis, especially for networks with heavy-tailed degree distributions typical of many real-world systems. In fact, they offer an enhanced visualization of the degree structure, reducing noise in the distribution's tail and facilitating the identification of long-tail or power-law behaviors indicative of scale-free networks.
    The distance between two cumulative distributions $P^{(c)}, \Tilde{P}^{(c)}$ can be quantified e.g. by computing the Kolmogorov–Smirnov statistic:
        \begin{equation}
            D_{KS} = \sup_k |P^{(c)}(k)-\Tilde{P}^{(c)}(k)|\ .
        \end{equation}
\end{itemize}

\section{A roadmap}
In this section, we point more in detail at relevant lines of development of our work.

\subsection{More detailed models of a worm brain}
The EE model of the \ce\ brain wiring dynamics designed in \ms\ lends itself to a number of extensions for achieving more detailed depictions of the underlying process. To start with, when formulating the biological function $F$, simplicity can be traded for realism in several ways.

In the first place, the system's representation. For instance, a natural extension of our approach would be to take into account the directed nature of chemical synapses, by considering directed dyads $(i\rightarrow j)$, hence directed graphs. From a technical point of view, this would entail storing $2L = N(N-1)$ bits per graph -- i.e., twice the memory $L$ needed for an undirected graph. The statistics eq.\eqref{e-stats} in \ms\ admit generalisations to the directed case, which however require non-trivial modelling choices, see \cite{krivitsky2022}. Directed graphs would allow modelling the existence of a feedforward bias for synaptogenesis that increases with time, i.e. new synaptic connections mainly appear in the direction: sensory neurons $\rightarrow$ (modulatory neurons $\rightarrow$) interneurons $\rightarrow$ motor neurons \cite{witvliet2021}. As usual, this should be specified by including a suitable term in the formulation of $F$. 

A general and useful ERG-like term to model effects is the edge covariate: 
\begin{equation}\label{e-ecov-d}
    x(G) = \frac{1}{2L}\sum_{i\ne j}a_{ij}\gamma_{ij}\ ,
\end{equation}
here in the case of directed graphs, cf eq.\eqref{e-edgecov}. The latter is a sum over all existing (directed) edges of the edge-attribute $\gamma_{ij}$. The latter can be used to specify a number of different effects. For this to be possible, the information about $\gamma_{ij}$ of each possible dyad must be available.
Three examples are the following.
\begin{itemize}
    \item[$\circ$] \emph{Feedworward bias}. We define $\gamma_{ij} = 1$ if the synaptic connections goes from sensory to motor neurons, $\gamma_{ij} = -1$ if it goes in the opposite direction, $\gamma_{ij} = 0$ if it connects neurons of the same time. Then, eq.\eqref{e-ecov-d} quantifies the feedforward bias.
    \item[$\circ$] \emph{Homophilies}. This refers to the fact that neurons in the worm's brain are more likely to form a connection if they share common attribute, see \cite{pathak2020} for an overview. For example, there is a bilaterally symmetric pairing homophily, i.e., neurons are more likely to be connected if they belong to a bilaterally symmetric pair, tab. \ref{t-neurons}. To account for such effect, we could use eq.\eqref{e-ecov-d} with $\gamma_{ij}=\delta({\alpha_{i},\alpha_{j}})$, where $\delta$ is the Kroenecker delta, and $\alpha_i = L,R$ if the $i$-th neuron belongs to the left or right part of the worm body, respectively.
    \item[$\circ$] \emph{Spatial embedding}. The worm brain network is embedded in the physical space \cite{nicosia2013}. Long connections are anticipated to incur higher penalties due to increased wiring costs. The simplest way to take this into account would be to use eq.\eqref{e-ecov-d} with $\gamma_{ij}$ corresponding to the euclidean distance between the cell bodies of each pair of neurons.
\end{itemize}

Furthermore, the biological function $F$ can be endowed with additional terms that can account for the presence of stable neuronal circuits in the wiring \cite{witvliet2021}. For example, suppose that the (directed) triangle $i \rightarrow j \rightarrow k \rightarrow i $ is of particular biological importance to the worm's nervous system and is therefore expected to be disproportionately conserved across individuals. A term of the form $x(G) = 1 - a_{ij}a_{jk}a_{ki}$ could be added to $F$. In the case of a negative associated parameter $\theta$, this term would reduce the likelihood of those graphs that do not have the above circuit. 

According to the spirit of the approach here illustrated, the idea would be to include these effects (some, all, more) in the formulation of the $F$ metric and let the ERG estimation "judge" which are the most relevant, as quantified by the associated parameter $\theta$. Large magnitudes would indicated relevancy, while $\theta\sim0$ would indicate marginal or no contributions to the observed network. More detailed models will allow for reproducing more detailed features of the data. A systematic exploration of the space of possible models along the lines outlined above would holds considerable interest and we envision this as a promising avenue for future investigations. 

\smallskip
Zooming out to the EE framework, several possible refinements could improve the model realism. A straightforward extension would implement the appearance of new neurons at different stages during the developmental dynamics, as mentioned in sec. \ref{sss-a_run}. A less straightforward generalisation would be to abandon the hypothesis of time homogeneity of the process. For instance, this would mean to use time dependent EE parameters $\mu(t), \rho(t)$, or even 
a time dependent functional landscape -- or, \emph{functional seascape} -- $F(G,t)$. Here we are currently limited by the availability of only a few snapshots of the worm brain, which limits the complexity of the hypotheses that can be meaningfully made and tested. However, given the rapid advances in experimental imaging techniques, this family of generalisations of the current approach may be within reach in the near future. 

\smallskip
Finally, if and when data will be available, our model could be readily extended to different kinds of brain connectivity, as for example the network of gap junctions within the \ce\ brain, or its functional connectivity \cite{randi2022}. Also, an exciting line of investigations would entail developing a similar model for different natural brains, whose reconstruction are currently at different levels of completion: larval zebrafish, \emph{Drosophila}, mouse (...) \cite{hildebrand2017,dorkenwald2023,lee2016}

\subsection{Beyond the brain wiring problem} 
The EE dynamics arise from the interplay between a stochastic exploration of the configuration space and a state-dependent exploitation of the more functional states, regardless of the data representation. It encodes the interplay between randomness and functional robustness of the dynamics. The brain wiring dynamics, we have argued, fits this theoretical picture. However, EE dynamics can be used to describe also other types of dynamics, that can be placed within the same conceptual framework. In the following, we anticipate some concrete possible examples (biological and non-biological) that could be studied under the EE lens.

\begin{itemize}
    \item[$\circ$] \emph{Human brain diseases}. Within connectomics applications, different questions can be asked, at different scales of neuronal organisation. EE dynamics could be used to understand how the human brain -- as reconstructed from electrophysiological or neuroimaging experiments -- longitudinally reorganises after a traumatic event. For example, during the recovery from stroke, the brain tends to spontaneously explore local rewiring strategies to reestablish functional connections both within the affected hemisphere and with the unaffected hemisphere \cite{siegel2018,dancause2005}. Although at a different scale of neural organisation - here nodes are brain regions, edges measure structural or functional connectivity -- a similar approach to that developed in \ms\ could be used to model individual recovery/reorganisation dynamics, with the same rationale.
    \item[$\circ$] \emph{Social contacts}. Social bonds between individuals are essential for many living organisms. Modeling the resulting network formation is therefore crucial to understand, predict, and eventually influence the overall system behavior. For example, the formation of a social network on digital platforms such as Twitter occurs when users are exposed to various potential connections, often through suggested posts in their feeds or recommended profiles. Some lead to the establishment of new social bonds, typically in the form of 'follow' relationships. The probability of this process is influenced by several factors, including the degree of homophily among users and the presence of mutual acquaintances  \cite{mislove2007}. When representing the social network as collections of nodes (users) and directed edges (indicating 'follow' relationships), these dynamics can be conceptualized as resulting from two key drivers: (1) a non-uniform, stochastic exploration of the graph space and (2) a state-dependent exploitation process driven by a ERG-like $F$ accounting for homophilies (edge covariates) and graph-topological effects -- e.g., the presence of hub nodes (popular individuals) and clustering (well-connected communities).
    \item[$\circ$] \emph{Spatial navigation}. Beyond the network representation, the EE paradigm could be used also to study the problem of spatial navigation in complex environments. For example, the slime mold \emph{Physarum polycephalum} uses a form of reactive navigation to explore its environment. In the presence of attractants like food, its cytoplasm flows stochastically toward the food source, while repellents like light or salts push it away. Moreover, as the slime mold forages, it leaves behind a trail of nonliving extracellular slime, which it later avoids. This simple, noneuronal organism is able to solve complex navigation tasks, such as the U-shaped trap problem \cite{reid2012}. In terms of an EE dynamics, this can be modeled as (i) a stochastic exploration of the physical space, and (ii) a state-dependent exploitation driven by a $F$ metric which enforces hard constraints (physical barrier and previous path), soft constraints (repellents) and favors those displacements that minimise the distance from the food source.
\end{itemize} 

\bibliography{mybib}